\documentclass{elsart}
\usepackage{graphicx}
\usepackage{amsmath}
\usepackage{amssymb}

\begin{document}

\newcommand {\ee}[1]{\label{#1}\end{equation}}
\newcommand{\be}{\begin{equation}}
\newcommand {\av}[1] {\left\langle #1 \right\rangle }
\def\reff#1{(\ref{#1})}
\def\re{\mbox{Re}}
\def\w{\omega}
\def\W{\Omega}
\def\t{\tau}
\def\e{\varepsilon}
\newcommand{\pd}{\partial}
\newcommand{\ds}{\displaystyle}
\newcommand{\fracpd}[2]{\displaystyle\frac{\partial #1}{\partial #2}}
\newcommand{\fracd}[2]{\displaystyle\frac{d #1}{d #2}}
\begin{frontmatter}

\title{Partially integrable dynamics of ensembles of 
nonidentical oscillators}
\author{Arkady Pikovsky and Michael Rosenblum}
\address{Department of Physics and Astronomy, University of Potsdam,
  Karl-Liebknecht-Str. 24/25, D-14476 Potsdam, Germany}
\date{\today}
\begin{abstract}%
We consider ensembles of sine-coupled phase oscillators 
consisting of subpopulations of identical units, 
with a general heterogeneous coupling between subpopulations. 
Using the Watanabe-Strogatz ansatz we 
reduce the dynamics of the ensemble to a relatively small number of 
dynamical variables plus microscopic constants of motion. This reduction is
independent of the sizes of subpopulations and remains valid in the
thermodynamic limits, where these sizes or/and the number of
subpopulations are infinite. 
We demonstrate that the approach to the dynamics of such systems,
recently proposed by Ott and Antonsen,  corresponds to a
particular choice of microscopic constants of motion.  
The theory is applied to the standard Kuramoto
model and to the description of two interacting subpopulations, exhibiting
a chimera state.
Furthermore, we analyze the dynamics of the extension of the Kuramoto 
model for the case of nonlinear coupling and demonstrate the multistability of 
synchronous states.
\end{abstract}
\today
\end{frontmatter}

\section{Introduction}

A model of all-to-all, or globally coupled limit cycle oscillators  explains
many natural phenomena in various branches of science. The applications 
range from the description of the collective dynamics of 
Josephson junctions \cite{Wiesenfeld-Swift-95}, lasers \cite{Glova-03}, and 
electrochemical oscillators \cite{Kiss-Zhai-Hudson-02a} to that of pedestrians
on footbridges \cite{Strogatz_et_al-05,Eckhardt_et_al-07}, applauding persons in a 
large audience \cite{Neda-Ravasz-Brechet-Vicsek-Barabasi-00}, 
cells, exhibiting glycolitic oscillations 
\cite{Richard-Bakker-Teusink-Van-Dam-Westerhoff-96,Dano-Sorensen-Hynne-99,Gonze-Markadieu-Goldbeter-08},
neuronal populations \cite{Golomb-Hansel-Mato-01}, etc.
Externally forced or feedback controlled globally coupled ensemble 
or several interacting ensembles serve as models of circadian rhythms, 
normal and pathological brain activity, interaction of different brain regions, 
and many other problems 
\cite{Sakaguchi-88,Antonsen-Faghih-Girvan-Ott-08,%
Tass-99,Rosenblum-Pikovsky-04b,Rosenblum-Pikovsky-04c,Childs-Strogatz-08,%
Martens_etal-09}.
Many aspects of the ensemble dynamics, especially those related to 
inhomogeneity 
\cite{Abrams-Mirollo-Strogatz-Wiley-08,Omelchenko-Maistrenko-Tass-08,Laing-09} 
or nonlinearity of coupling \cite{Rosenblum-Pikovsky-07,Pikovsky-Rosenblum-09}, 
temporal dynamics of the collective mode \cite{Ott-Antonsen-08,Ott-Antonsen-09}, 
different frequency distributions \cite{Pazo-05,Pazo-Montbrio-09},
and clustering \cite{Kori-Kuramoto-01,Liu-Lai-Hoppensteadt-01,%
Maistrenko-Popovych-Burylko-Tass-04} remain in the focus of the current research activity. 

Ensembles of weakly interacting units are successfully treated within the 
framework of phase approximation 
\cite{Kuramoto-75,Kuramoto-84,Daido-92a,Daido-93a,Daido-96}.
Most popular is the Kuramoto model of sine-coupled phase oscillators.
This model explains self-synchroni\-za\-tion and appearance of a collective mode 
(mean field) in
an ensemble of generally non-identical elements; the transition to synchrony occurs 
at a certain critical value of the coupling constant that is roughly proportional 
to the width of the distribution of natural frequencies \cite{Kuramoto-75,Kuramoto-84}.
With the further increase of coupling, more and more oscillators join synchronous 
cluster, so that the amplitude of the mean field grows as a square root of 
supercriticality. 
It is instructive to interprete this transition as follows: the non-zero mean field 
forces individual units and entrains at least a part of them; these entrained units
become coherent, thus yielding a non-zero mean field. A quantitative consideration, based
on this argument and first performed by Kuramoto \cite{Kuramoto-75,Kuramoto-84}, provides
the amplitude and frequency of the stationary solution.
References to many further aspects of the
Kuramoto model can be found
in~\cite{Pikovsky-Rosenblum-Kurths-01,Acebron-etal-05,Strogatz-00}.

An extension of the Kuramoto model for the case of nonlinear coupling has been suggested
in our recent publications \cite{Rosenblum-Pikovsky-07,Pikovsky-Rosenblum-09}, where the most 
simple case of identical units has been treated. 
Nonlinearity in this context  means that the effect of the collective mode on an individual unit
depends on the amplitude of this forcing, so that, e.g., the interaction of the field and a unit  
can be attractive 
for weak forcing and repulsive for strong one. This can lead to nontrivial effects like a 
destruction of a completely synchronous state and appearance of partial synchrony in an 
ensemble of identical units. Moreover, in this state the frequencies of the collective mode 
and of oscillators can be different and incommensurate. 
The analysis of the nonlinear model for the case of an ensemble with a frequency distribution 
is still lacking and will be performed below. 
This analysis is based on the extension of the Watanabe--Strogatz (WS) 
theory~\cite{Watanabe-Strogatz-93,Watanabe-Strogatz-94} to the case of nonidentical 
oscillators, suggested in our brief communication \cite{Pikovsky-Rosenblum-08}. 

For a population of \textit{identical} oscillators, a full dynamical description of the 
Kuramoto model can be obtained with the help of the WS ansatz
which reduces the dynamics to that of three macroscopic variables plus constants of motion. 
This works even for a nonlinear coupling \cite{Pikovsky-Rosenblum-09}.
The main idea of this paper is in 
extending the WS ansatz on an ensemble of 
nonidentical oscillators, treating it 
as a system of coupled subpopulations, each consisting of 
identical oscillators. 
Each subpopulation can be then described by three WS variables, whereas the full system 
is described by a system of coupled WS equations. 
A description of an ensemble with a continuous frequency distribution is then obtained 
by performing a thermodynamic limit.
The extended WS approach is applied to a population of oscillators with a 
Lorentzian frequency distribution (the standard Kuramoto model), and to a population of identical 
oscillators with inhomogeneous coupling \cite{Abrams-Mirollo-Strogatz-Wiley-08}.
In particular, within this framework we establish a relation between the WS approach and the
recent Ott-Antonsen (OA) ansatz \cite{Ott-Antonsen-08,Ott-Antonsen-09}, 
which yields, under certain 
assumptions, a dynamical equation for the evolution of the mean field.
In this paper we discuss in detail how the WS theory can be applied to populations of 
non-identical units and apply this approach to describe the dynamics of 
oscillator ensembles with global nonlinear coupling, for the case of 
Lorentzian  distribution of frequencies.

The paper is organized as follows. We discuss the main model in Section~\ref{secmod}.
Extension of the WS theory to the case of nonidentical oscillators is presented in 
Section~\ref{ws}. In Section~\ref{OAansatz} we discuss a relation between the WS theory and the OA ansatz (cf. a recent paper
by Marvel,  Mirollo, and Strogatz~\cite{Marvel-Mirollo-Strogatz-09} for a related discussion). 
This theory is applied to describe the dynamics of linearly 
(Section~\ref{sec:lin}) and nonlinearly (Section~\ref{sec:nonlin})
coupled ensembles; here we also support the theory by numerics.
In particular, we demonstrate the differences of the dynamics of the full equation system
and that confined to the OA reduced manifold.
We summarize and discuss our results in Section~\ref{conc}.

\section{Hierarchically organized population of oscillators}
\label{secmod}

In this Section we introduce a model of hierarchically organized populations of 
oscillators and describe it in terms of microscopic equations of motion. 
The main idea is to treat an ensemble of nonidentical oscillators as a 
collection of subpopulations of identical oscillators.

We start by introducing 
an ensemble of nonidentical phase oscillators, characterized by phase 
variables $\phi_k$ and generally time-dependent frequencies $\w_k(t)$, 
where $k=1,\ldots,N$ is the oscillator index and $N$ is the ensemble size.
Each oscillator generally interacts with all other units and is subject 
to external fields. The effect of these interactions can be represented as
some effective forcing and therefore each unit is described as a driven 
oscillator
 \begin{equation}
\fracd{\phi_k}{t}=\w_k(t)+A_k(t)\sin(\xi_k(t)-\phi_k)\;,\quad k=1,\ldots,N\;,
\label{hopo-1} 
\end{equation}
where variables $A_k$ and $\xi_k$ characterize the amplitude and phase of the force. 
It is convenient to introduce a complex force $H_k=A_ke^{i\xi_k}$ and re-write 
Eq.~\reff{hopo-1} as
\begin{equation}
 \fracd{\phi_k}{t}=\w_k+\mbox{Im}\left (H_k e^{-i\phi_k}  \right)\;.
\label{hopo-2}
\end{equation}

In many particular cases considered below the force $H$ is calculated in some 
mean-field manner from the state of the whole population or some subpopulation(s).
However, for a while we prefer to consider $H$, as well as $\w$, as arbitrary functions 
of time; for brevity we skip this dependence in the notations.
Note that generally $H$ can include random component(s) which describe a
forcing by common noise.
We emphasize that Eqs.~\reff{hopo-2} do not represent the most general model of 
coupled phase oscillators, because the high order harmonic components
of the phase $\sim e^{-in\phi_k}$, with $n>1$, do not appear on the right hand side 
of Eqs.~\reff{hopo-2}.

\begin{figure}[ht!]
\centerline{\includegraphics[width=0.49\textwidth]{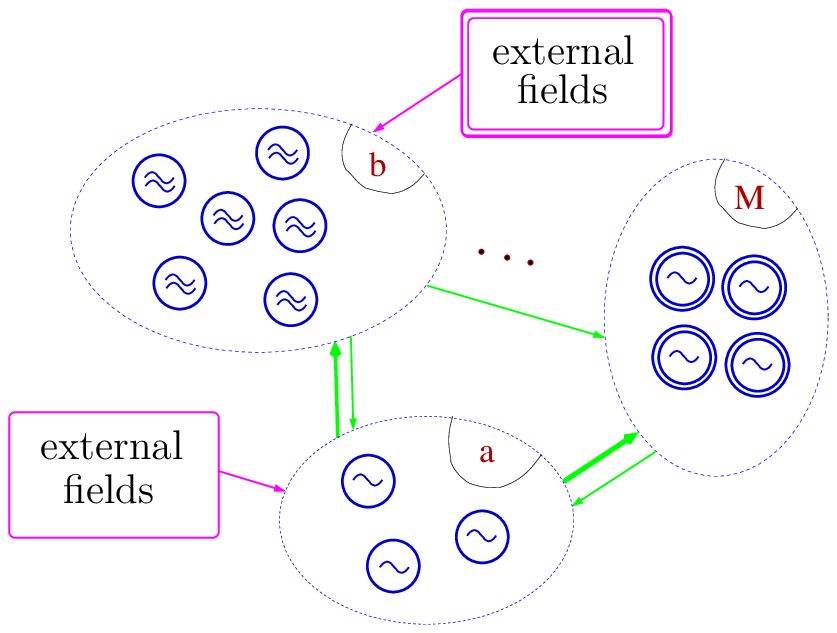}
\hfill
\includegraphics[width=0.49\textwidth]{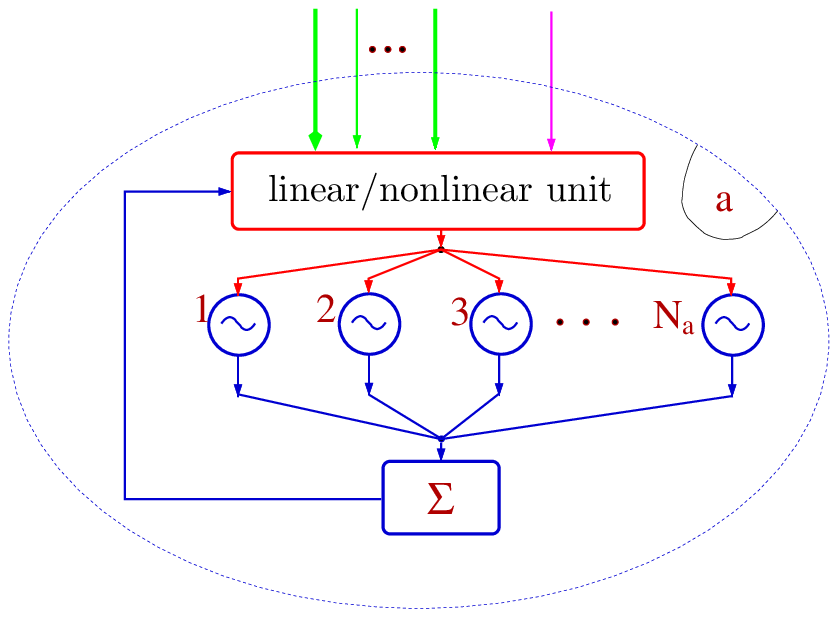}}
\caption{Hierarchically organized ensemble (left panel) can be considered as a collection of 
subpopulations labeled by index $a=1,\ldots,M$, with a generally bidirectional 
asymmetrical coupling between subpopulations.
Generally, the subpopulations are exposed to external field(s).
Each subpopulation (right panel) of the oscillator ensemble consists of identical oscillators 
driven  by a common force. This force results from the action of all other subpopulations and 
external fields and from the coupling within the subpopulation, which is usually 
taken to be of a mean field type. This means that outputs of all oscillators create the mean 
field that acts back on these oscillators. The crucial point is 
that the forcing 
is formed either via linear or nonlinear transformation of acting mean
fields.
}
\label{popstructure}
\end{figure}

In general, all oscillators described by \reff{hopo-2} have different
dynamics, and such a system cannot be further reduced. Such a
simplification is possible for an ensemble of {\em identical}
oscillators, where the WS theory holds
which states that dynamics of a population of 
identical oscillators of type \reff{hopo-2} subject to a
common forcing is effectively three-dimensional 
(this is discussed in details in 
the next Section).
In order to make use of the WS ansatz to an ensemble of \textit{nonidentical} 
oscillators, we assume that this ensemble consists of groups
(subpopulations) of identical units (Fig.~\ref{popstructure}, cf.~\cite{Barreto-Hunt-Ott-So-08}). 
Accordingly, 
we re-label the oscillators grouping them into $M$ subpopulations, 
denoted by index $a$.
Each population contains $N_a$ identical units with frequency $\w_a$ which are driven 
by a common force $H_a$.  The equations now read:
\begin{equation}
 \fracd{\phi_k^{(a)}}{t}=\w_a+\mbox{Im}\left (H_a e^{-i\phi_k^{(a)}}  \right)\;,
\label{hopo-3}
\end{equation}
where $k=1,\ldots,N_a$ and $a=1,\ldots,M$, with an obvious relation 
$\sum_{a=1}^M N_a=N$. 
Note that two subpopulations can have the same frequencies $\w_a=\w_b$ but differ 
by the force $H_a\ne H_b$, or they can have different frequencies but the same
force, or they can differ both by their frequencies and the forces.

Certainly, such grouping of oscillators into subpopulations is not always possible
(if we exclude a trivial case when each subpopulation contains one unit and $M=N$). 
However, in many particular cases $M\ll N$ and the description of the dynamics simplifies 
essentially: as we will see in the next Section, 
each group is described by three WS variables and instead of $N$ 
equations (\ref{hopo-3}) we have to deal with $M$ coupled systems of three WS equations each. 
Note that WS ansatz is also valid for \textit{infinitely large} populations of identical
oscillators. Thus, if some or all of subpopulation sizes $N_a\to\infty$, we still obtain a
$3M$-dimensional description of the collective dynamics.
As discussed in the next Section, the idea to consider an ensemble as a collection 
of subpopulations is also fruitful for treating a thermodynamic limit $N\to\infty$
with a \textit{continuous distribution} of oscillator frequencies.

\section{Oscillator populations in external fields}
\label{ws}

In this Section we discuss the dynamics of large populations of oscillators subject 
to an arbitrary external force.
First, we briefly present the WS theory which treats ensembles of identical oscillators. 
We re-write the WS equations in new notations, what makes them more convenient 
for the consequent analysis.
Next, we extend the WS theory to the case of a finite number of
interacting groups.
Finally, performing a thermodynamic limit, we obtain a description of an ensemble 
with a continuous frequency distribution. 

\subsection{Identical oscillators}
First we discuss in details dynamics of a population of identical oscillators subject to 
an arbitrary forcing, which is, however, common for all oscillators.

\subsubsection{WS equations}
The main result of the seminal papers by Watanabe and Strogatz 
\cite{Watanabe-Strogatz-93,Watanabe-Strogatz-94}
is that a population of $N>3$ sine-coupled phase oscillators with any time-dependent
common frequency $\w(t)$, driven by an arbitrary but common force $H(t)$, 
\be
\fracd{\phi_k}{t}=\w(t)+\mbox{Im}\left (H(t) e^{-i\phi_k}  \right)\;, \quad k=1,\ldots,N\;,
\ee{kur5}
 can be completely described by three global variables  $\rho(t)$, $\Phi(t)$, and $\Psi(t)$
plus constants of motion  $\psi_k$ that are determined by initial 
conditions.\footnote{There is a restriction: 
an initial state of the ensemble 
cannot contain too large clusters, see~\cite{Watanabe-Strogatz-94} for
details.}
The original phases can be recovered by means of the time-dependent transformation 
\begin{equation}
 e^{i\phi_k}=e^{i\Phi}\frac{\rho+ e^{i(\psi_k-\Psi)}}{\rho e^{i(\psi_k-\Psi)}+1}\;,
\label{ws-trans}
\end{equation}
or, equivalently,
\begin{equation}
\tan \left( \frac{\phi_k-\Phi}{2}\right) =\frac{1-\rho}{1+\rho}\tan \left( \frac{\psi_k-\Psi}{2}\right) \;,
\label{ws-trans1}
\end{equation}
see Fig.~\ref{transf} for illustration.\footnote
{
Note that transformation $\psi\to\phi$ has a form of the M\"obius transformation 
\cite{Marvel-Mirollo-Strogatz-09}.
}

Here $\psi_k$ are the constants of motion. Notice that only $N-3$ of them are independent;
see the discussion below. The time-dependent functions $\rho(t)$, $\Phi(t)$, and $\Psi(t)$ 
are the global amplitude and phase variables, respectively. 
In the following we denote them as WS variables, their meaning is discussed later. 
For the transformation \reff{ws-trans} to be consistent with equations
of motion \reff{kur5}, these variables have to obey the WS equations
\begin{align}
\frac{d\rho}{dt}&=\frac{1-\rho^2}{2}\mbox{Re}(He^{-i\Phi})\;,          \label{ws-1}\\[1ex]
\frac{d\Phi}{dt}&=\w+\frac{1+\rho^2}{2\rho}\mbox{Im}(He^{-i\Phi})\;,\label{ws-2}\\[1ex]
\frac{d\Psi}{dt}&=\frac{1-\rho^2}{2\rho}\mbox{Im}(He^{-i\Phi})\;.      \label{ws-3}
\end{align}
We emphasize that we use the global variables slightly different from those originally used by WS 
\cite{Watanabe-Strogatz-94} and write the equations in a different form. 
The relation to the original form can be found in Appendix~\ref{AppWS}; 
there we also demonstrate the equivalence of the transformations (\ref{ws-trans}) and 
(\ref{ws-trans1}).

\begin{figure}[ht!]
\centerline{\includegraphics[width=0.4\textwidth]{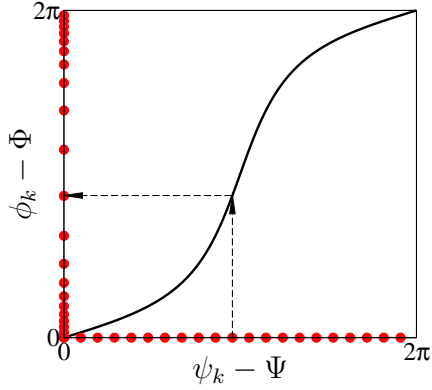}}
\caption{Illustration of the transformation $\psi_k\to\phi_k$, 
see Eqs.~(\ref{ws-trans},\ref{ws-trans1}), here for $\rho=0.5$
(cf. Fig.~3 in \protect\cite{Watanabe-Strogatz-94}).
}
\label{transf}
\end{figure}

For a further analysis it is convenient to introduce a combination of two WS variables
$z=\rho e^{i\Phi}$. We call this complex variable the bunch amplitude. 
Introducing  also the 
phase shift $\alpha=\Phi-\Psi$, we write the WS equations (\ref{ws-1}-\ref{ws-3})
in an equivalent form:\footnote{We note here that Eq.~(\ref{ws-4}) coincides with the Ott-Antonsen equation 
\cite{Ott-Antonsen-08} for the dynamics of the complex mean field. However, in the 
Ott-Antonsen ansatz it appears without Eq.~(\ref{ws-5}). 
The relation between the OA ansatz and the WS theory is treated in details 
in Section~\ref{OAansatz}.}
\begin{align}
\frac{dz}{dt}&=i\w z+\frac{1}{2}H-\frac{z^2}{2}H^*\;, \label{ws-4}\\[1ex]
\frac{d\alpha}{dt}&=\w+\text{Im}(z^*H)\;.\label{ws-5}
\end{align}

\subsubsection{Constants of motion}
Transformation (\ref{ws-trans}) from original variables $\phi_k$ to WS variables 
$\rho$, $\Phi$, $\Psi$ (or $z$ and $\alpha$)
and $\psi_k$ yields an overdetermined system. Hence, we have to impose $3$ constraints 
on the constants $\psi_k$; this is discussed in detail in \cite{Watanabe-Strogatz-94}.
It is convenient to choose two constraints as follows:
\begin{equation}
 \sum_{k=1}^N e^{i\psi_k}=0\;.
\label{psicond}
\end{equation}
The third constraint on $\psi_k$ is somehow arbitrary, it only fixes
the common shift of $\psi_k$ with respect to $\Psi$. 
It can be taken, e.g., as $\sum_k \psi_k=0$ 
(this implies that constants are defined 
in the $[-\pi,\pi]$ interval) \cite{Watanabe-Strogatz-94}.
Another convenient choice is $\sum \cos (2\psi_k)=0$.
The imposed constraints allow one to determine unambiguously the 
new variables $\rho(0)$, $\Psi(0)$, $\Phi(0)$ and constants $\psi_k$ 
from the initial conditions $\phi_k(0)$ and vice versa, this is
discussed in details in Ref.~\cite{Watanabe-Strogatz-94}.

\subsubsection{Meaning of the WS variables}
In order to discuss the physical meaning of the WS variables we compare
them with the 
complex mean field, or the Kuramoto order parameter
\begin{equation}
 Z=N^{-1}\sum_{k=1}^N e^{i\phi_k}=re^{i\Theta},
\label{orderpar}
\end{equation}
where $r$ and $\Theta$ are the amplitude and phase of the mean field, respectively.
[The comparison here is qualitative; a quantitative relation is given in 
Section~\ref{OAansatz}.]

The WS amplitude variable $\rho$ is roughly 
proportional to the mean field amplitude $r$. Indeed, if $\rho=0$, then from Eq.~(\ref{ws-trans}) 
with account of Eq.~(\ref{psicond}), we obtain $r=0$. 
Similarly, Eq.~(\ref{ws-trans}) shows that for $\rho=1$ all $\phi_k$ are equal, what yields $r=1$.
For intermediate values $0<\rho<1$ the relation between $\rho$ and 
$r$ generally depends also on $\Psi$ and $\psi_k$. 

The WS phase variable $\Phi$ characterizes the position of the 
maximum in the distribution of phases and is close to the phase of the mean 
field $\Theta$. They coincide for $\rho=r=1$; for $0<\rho<1$, $\Phi$ is shifted with respect 
to $\Theta$ by a factor, dependent on $\rho$, $\Psi$.

\begin{figure}[ht!]
\centerline{\includegraphics[width=0.4\textwidth]{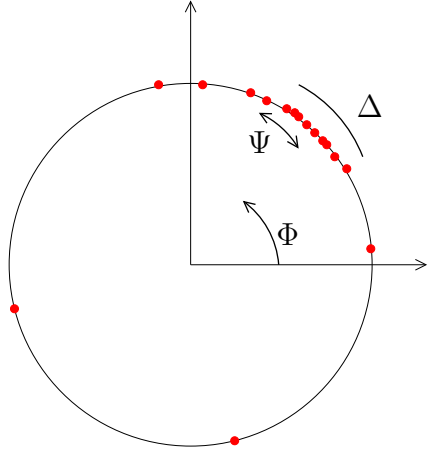}}
\caption{Illustration of the meaning of the WS variables. 
The global amplitude $\rho$ is related to the width $\Delta$ of the distribution 
of phases; $\rho=0$ if this distribution is uniform and $\rho=1$ if it collapses 
to a $\delta$-distribution. Thus, $\rho$ is roughly proportional to the mean field 
amplitude $r$. Angle variable $\Phi$ describes the position of the hump in the 
distribution; therefore it roughly corresponds to the phase $\Theta$ of the 
mean field. Angle variable $\Psi$ characterizes the motion of individual oscillators 
with respect to the hump.
}
\label{interprws}
\end{figure}

The second WS phase variable $\Psi$ determines the shift of individual oscillators
with respect to $\Phi$. 
From Eqs.~(\ref{ws-2},\ref{ws-3}) we obtain a useful relation:
\begin{equation}
 \dot\Psi=\frac{1-\rho^2}{1+\rho^2}(\dot\Phi-\w)\;.
\label{dPsi_dPhi}
\end{equation}
As can be seen from Fig.~\ref{transf} and Eq.~(\ref{ws-trans1}),
$\phi_k$ decreases by $2\pi$ if $\Psi$ grows by $2\pi$.
Hence, the oscillator frequency is
\begin{equation}
 \w_{osc}=\langle \dot\phi_k\rangle =\langle\dot\Phi\rangle-\langle\dot\Psi\rangle \;.
\label{oscfreq}
\end{equation}

\subsubsection{Dynamics of WS variables}
\label{dws}
Here we discuss possible solutions of Eqs.~(\ref{ws-1}-\ref{ws-3}) for given $\w(t)$, $H(t)$.
First, we note that these equations represent a skew system: 
the variable $\Psi$ does not enter Eqs.~(\ref{ws-1},\ref{ws-2}). 
Hence, one has to solve the two-dimensional system (\ref{ws-1},\ref{ws-2}) and then obtain 
$\Psi(t)$ via Eq.~\reff{dPsi_dPhi}.
Next, the system~(\ref{ws-1}-\ref{ws-3}) obviously admits a fully
synchronous solution $\rho=1$,
describing a synchronized cluster of identical oscillators: all in the
same state.
As it follows from Eq.~(\ref{ws-3}), in this case $\Psi$ is an arbitrary constant and the global 
dynamics is described solely by an equation for $\Phi$. 
However, we cannot analyze stability of this solution for general functions $\w(t)$, $H(t)$.
We mention here a seemingly trivial state when $H_0=0$.  
Then $\rho$ is arbitrary, $\dot\Phi=\w$, and $\Psi=const$. This solution appears in the 
nonlinear model treated below in Section~\ref{sec:nonlin}.

It is instructive to find solutions of Eqs.~(\ref{ws-1}-\ref{ws-3}) for an important particular 
case $\w=\mbox{const}$ and $H(t)=H_0e^{i\nu t}$. For this case of a
harmonic forcing, Eqs.~(\ref{ws-1}-\ref{ws-3}) 
take the form:
\begin{align}
\frac{d\rho}{dt}&=\frac{1-\rho^2}{2}H_0\cos(\nu t-\Phi)\;,          \label{ws1-1}\\[1ex]
\frac{d\Phi}{dt}&=\w+\frac{1+\rho^2}{2\rho}H_0\sin(\nu t-\Phi)\;,     \label{ws1-2}\\[1ex]
\frac{d\Psi}{dt}&=\frac{1-\rho^2}{2\rho}H_0\sin(\nu t-\Phi)\;.     \label{ws1-3}
\end{align}
Introducing the phase differences $\Delta=\ds\frac{\pi}{2}+\Phi-\nu t$ and
$\alpha=\Phi-\Psi$
we rewrite this system as an autonomous one
\begin{align}
\frac{d\rho}{dt}&=\frac{1-\rho^2}{2}H_0\sin\Delta\;,          \label{ws2-1}\\[1ex]
\frac{d\Delta}{dt}&=\w-\nu+\frac{1+\rho^2}{2\rho}H_0\cos\Delta\;,    \label{ws2-2}\\[1ex]
\frac{d\alpha}{dt}&=\w-\nu-H_0\rho\cos\Delta\;. \label{ws2-3}
\end{align}

Remarkably, the system (\ref{ws2-1},\ref{ws2-2}) is reversible, 
as it remains invariant under a transformation $\Delta\to -\Delta$, $t\to -t$.
The system possesses two types of solutions, in dependence on the detuning $|\w -\nu|$.
For $|\w-\nu|\le H_0$  system (\ref{ws2-1},\ref{ws2-2})
has one attractive and one repelling steady state solution:
these fixed points are located symmetrically with respect to the
involution $\Delta\to -\Delta$ and have coordinates
$\rho=1$ and $\Delta=\pm\arccos \frac{|\w-\nu|}{H_0}$, 
respectively. 
For the case $|\w-\nu|>H_0$ the system has no attractors, and the exact
solution, 
described in Appendix~\ref{AppWE}, 
is represented by a family of closed orbits around a marginally stable equilibrium 
with coordinates $\Delta_0=0$ and $\rho_0$ (see Eq.~\reff{fixp}).
For this equilibrium solution, $\Psi$ rotates with
a constant frequency, given by Eq.~(\ref{fixpfreq}).
Except for this special case, the variables $\rho$ and $\Delta$ also 
oscillate in time; moreover, the variable $\Psi$ possesses an additional frequency 
so that the full dynamics is quasiperiodic (see Appendix~\ref{AppWE}).

In summary, the behavior of the ensemble of non-interacting oscillators 
in a common field ``follows'' the dynamics of an individual oscillator. 
If the latter is entrained by the force then the ensemble is also fully 
synchronized; this corresponds to an attractive fixed point solution of
system (\ref{ws2-1},\ref{ws2-2}). 
If an individual oscillator is not locked to the field, 
then there is no effective dissipation in the ensemble and system (\ref{ws2-1},\ref{ws2-2})
has no attractive solutions; as a result the collective variables oscillate. 
Only for a special preparation of initial conditions these variables are $\rho=\rho_0$ 
and $\Delta=0$, i.e. they do not vary in time. 
This dynamics of the collective variables has a clear physical meaning if we 
interprete it as a dynamics of macroscopic characteristics of the population distribution.
If the oscillators are locked, the population distribution collapses to a $\delta$-function,
what corresponds to $\rho=1$, and the only relevant quantity is the position 
of the cluster $\Phi$. 
If the oscillators are not locked, then each of them rotates non-uniformly with 
respect to the external field, and the distribution is generally ``breathing'' 
-- unless the initial conditions correspond to the steady distribution, i.e. 
to the marginal equilibrium point $\rho=\rho_0$ and $\Delta=0$.  
From this picture follows an important fact: the time averages along 
all possible trajectories of (\ref{ws2-1}-\ref{ws2-3}) are equal.  
Indeed, average of a macroscopic variable can be obtained by averaging over time average 
for individual oscillators, and the latter quantities are the same because 
the oscillators are identical.

\subsection{Ensemble of nonidentical units as a hierarchical population}

Consider now an ensemble which can be viewed at as a collection of $M$
subpopulations of identical units. Let index $a=1,\ldots,M$ label the 
subpopulations and let each subpopulation consist of $N_a>3$ oscillators. 
Again, we assume that the force $H_a$ acting on a subpopulation $a$ equally 
effects all oscillators of this subpopulation.
Hence, the dynamics of each subpopulation can be described by means of the 
WS ansatz.
For the $a$-th subpopulation we write, similar to (\ref{ws-4},\ref{ws-5})
\begin{align}
\frac{dz_a}{dt}&=i\w_a z_a+\frac{1}{2}H_a-\frac{z_a^2}{2}H_a^*\;, \label{hp-1}\\[1ex]
\frac{d\alpha_a}{dt}&=\w_a+\text{Im}(z_a^*H_a)\;.\label{hp-2}
\end{align}
For each subpopulation we have to specify constants of motion $\psi_{a,k}$, 
$k=1,\ldots,N_a$; each set of these constant obeying 
the same three additional constraints, see Eq.~(\ref{psicond}) and the paragraph 
following it.

Hence, the full hierarchical system is described by $2M$ equations for $M$ complex 
variables $z_a$ and $M$ real variables $\alpha_a$, plus $N-3M$ constants of motion.
(Certainly, we can use the WS equations in the real form (\ref{ws-1}-\ref{ws-3}) 
to obtain $3M$ equations for $3$ real variables.)

\subsubsection{Treating individual oscillators and clusters}
Having introduced the coupled WS equations for several interacting subpopulations
we come back to  the limitation of the WS theory: the number of
oscillators should be larger than three, and an initial configuration of a 
subpopulation cannot have too large clusters of fully identical
oscillators. 
To overcome this, it is sufficient to notice that Eq.~\reff{hopo-2}
for an individual oscillator can be also written in form
(\ref{ws-1}-\ref{ws-3}) if we set $\rho=1$, $\Phi=\phi$, and $\Psi$
as an arbitrary constant. In this case in \reff{hp-1} we have $|z|=1$ and the
two equations of system (\ref{hp-1},\ref{hp-2}) are in fact equivalent.
Thus, individual oscillators not belonging to large groups can be
also described by Eqs.~(\ref{hp-1},\ref{hp-2}).

The same idea helps to treat large clusters in subpopulations.
Suppose that a subpopulation $a$ with $N_a$ elements does have a fully synchronized 
cluster formed by $N_{a}^{(c)}>N_a/3$ oscillators.
To treat such a group, we split this subpopulation into two, labeled by 
$a'$ and $a''$, respectively. 
(It means that we have now $M+1$ subpopulations.)
The first one is formed by $N_a^{(c)}$ elements of the cluster, the second one 
is formed by other $N_a-N_a^{(c)}$ oscillators.
The first subpopulation can be treated like one oscillator: 
it  has $\rho_{a'}=1$ and is described by one equation for the phase 
$\Phi_{a'}$ (see Section~\ref{dws}); the second one is described by three WS equations, 
so that altogether we have four  equations for this subpopulation.
If it turns out that the population $a''$ also contains a majority cluster, then again 
the elements of this cluster can be considered as a separate subpopulation, and so on.

\subsection{Infinitely large populations}
Most applications of the theory treat infinitely large ensembles.
Here we consider separately two cases: (i) the number of subpopulations remains finite, 
but the subpopulations are infinitely large and (ii) the number of subpopulations is infinite.

\subsubsection{Thermodynamic limit I: {F}inite number of subpopulations}
\label{tl1}

If all or some of $M$ subpopulations are infinitely large, they still can be described 
by Eqs.~(\ref{hp-1},\ref{hp-2}). However, any infinitely large subpopulation is now described not 
by a finite set of constants of motion $\psi_{a,k}$, but by the distribution 
functions $\sigma_a(\psi)$ (see~\cite{Watanabe-Strogatz-94}). Equations~(\ref{psicond})
take the form
\begin{equation}
 \int_{-\pi}^{\pi}\sigma_a(\psi) e^{i\psi}d\psi=0\;.
\label{psicond1}
\end{equation}
The third constraint for the constants of motion is also written via an integral, 
e.g., $\int_{-\pi}^{\pi}\sigma(\psi)\cos 2\psi d\psi=0$.
Correspondingly, the phases are described by a distribution density
$w_a(\phi)$.

\subsubsection{Thermodynamical limit II: {I}nfinite number of subpopulations}

Now we assume that the number of subpopulations $M\to\infty$.
Hence, we substitute the subpopulation index $a$  by a continuous variable, say $x$. 
In most applications this variable will be associated with the frequency, and in 
this way we will be able to describe ensembles with a continuous distribution of 
frequencies, however at the moment we prefer not to specify the meaning of $x$.
(Generally, $x$ can be also a vector variable.)

Thus, the WS variables $z$, $\alpha$ or $\rho$, $\Phi$, $\Psi$, as well as the forcing $H$ 
become functions of $x$ and $t$ and the WS equation system takes the form:
\begin{align}
\frac{\partial z(x,t)}{\partial t}&=i\w(x,t) z+\frac{1}{2}H(x,t)-\frac{z^2}{2}H^*(x,t)\;, 
\label{ws-c1}\\[1ex]
\frac{\partial\alpha(x,t)}{\partial t}&=\w(x,t)+\text{Im}\left ( z^*H(x,t)\right )\;.
\label{ws-c2}
\end{align}
In performing this limit we can consider the subpopulations as finite or infinite. 

We note that generally, the description of the system in this limit 
remains infinitely dimensional. However, it is simpler than the original one 
because for each value of the continuous variable $x$, which can have, e.g., 
the meaning of frequency, we have only three WS variables.
Moreover, as is discussed below, at least for the case of global coupling with the 
Lorentzian frequency distribution, the description becomes really low-dimensional.

\subsubsection{Direct WS reduction for a system with continuous distribution of parameters}

Above we have treated an ensemble with a continuous distribution of
parameters as a thermodynamic limit of a hierarchically organized
populations. As the number of elements at each value of the continuous
parameter is arbitrary, it appears instructive to apply the WS ansatz
directly to an equation for the distribution density. Our derivation, presented
below, is heavily based on the derivation given by WS 
(see Ref.~\cite{Watanabe-Strogatz-94}), where they treated the case of 
identical oscillators.

We start with the continuity equation which expresses the conservation of the 
number of oscillators:
\begin{equation}
 \fracpd{w}{t}+\fracpd{}{\phi}(wv)=0\;,
\label{conteq}
\end{equation}
where $w(x,\phi,t)$ is the distribution density function. The velocity  $v=\dot\phi$ is 
determined by the microscopic equation of motion
\[
 v=\w(x,t)+\mbox{Im}\left[H(x,t)e^{i\phi}\right]\;.
\]
Following the idea of Watanabe and Strogatz \cite{Watanabe-Strogatz-94} we demonstrate 
that, with the transformation to the WS variables $\rho$, $\Phi$, $\Psi$ and $\psi$, the 
time-dependent density $w(x,\phi,t)$ is transformed into a stationary
density $\sigma(x,\psi)$.

We perform the following variable substitution in the continuity equation:
\[
t,\;\phi,\;x\quad\to\quad\tau=t,\;\psi=\psi(x,\phi,t),\;y=x \;.
\]
The relation between the densities in old and new variables takes the form:
\begin{equation}
 w(x,\phi,t)=\sigma(y,\psi,\tau)\fracpd{(y,\psi,\tau)}{(x,\phi,t)}= 
\sigma(x,\psi,\tau)\fracpd{\psi}{\phi} \;.
\label{eqdensnew}
\end{equation}
Writing the continuity equation in new coordinates (see Appendix~\ref{AppCE}), we obtain:
\begin{equation}
 \begin{array}{ll}
 0=&\fracpd{w}{t}+\frac{\pd}{\pd \phi}(w v) =
\fracpd{\sigma}{\tau}\fracpd{\psi}{\phi}+\fracpd{\sigma}{\psi}\left\lbrace 
\fracpd{\psi}{\phi}\left (\fracpd{\psi}{t}+v\fracpd{\psi}{\phi}\right)\right\rbrace\\[4ex]
&+\sigma\left\lbrace \ds\frac{\pd}{\pd \tau}\left( \fracpd{\psi}{\phi}\right) 
+\ds\frac{\pd}{\pd \psi} \left( \fracpd{\psi}{\phi}\right)
\left (\fracpd{\psi}{t}+v\fracpd{\psi}{\phi}\right )+\left (\fracpd{\psi}{\phi}\right) ^2\fracpd{v}{\psi}
\right\rbrace\;.
\end{array}
\label{cenew}
\end{equation}
It Appendix~\ref{AppCE} we show that both expressions in curly brackets vanish 
provided $\Phi(x,t)$, $\Psi(x,t)$, $\rho(x,t)$ obey :
\begin{equation}
\begin{array}{ll}
\ds\fracpd{\rho(x,t)}{t}&=\ds\frac{1-\rho^2}{2}\mbox{Re}(H(x,t)e^{-i\Phi})\;,\\[3ex]
\ds\fracpd{\Phi(x,t)}{t}&=\w(x,t)+\displaystyle\frac{1+\rho^2}{2\rho}\mbox{Im}(H(x,t)e^{-i\Phi})
\;,\\[3ex]
\ds\fracpd{\Psi(x,t)}{t}&=\ds\frac{1-\rho^2}{2\rho}\mbox{Im}(H(x,t)e^{-i\Phi})\;.
\end{array}
\label{wsnce}
\end{equation}
This implies that $\fracpd{\sigma}{\tau}=0$ and, thus, $\sigma(x,\psi)$ is a stationary distribution. 
Hence, the transformation to WS variables indeed results in a low-dimensional description 
of the dynamics (three global variables and function $\sigma$ for each value of $x$).
Equations (\ref{wsnce}) are equivalent to Eqs.~(\ref{ws-c1},\ref{ws-c2}).

\section{Linking the Watanabe-Strogatz theory and the Ott-Antonsen ansatz}
\label{OAansatz}
In this Section we relate WS variables to the complex mean field (which is the
Kuramoto order parameter, 
see Eq.~(\ref{orderpar})), and to the generalized Daido order parameters. 
We demonstrate that particular solutions of the WS equations for 
the uniform distribution of constants of motion are equivalent 
to the solutions obtained with the help of the Ott and Antonsen ansatz
\cite{Ott-Antonsen-08,Ott-Antonsen-09}, see also 
\cite{Lee-Ott-Antonsen-09,Abdulrehem-Ott-09}
Next, we discuss properties of the OA solution for singular and continuous 
distributions of oscillator frequencies. Note, that a relation between the WS and OA
theories has been also recently discussed by Marvel, Mirollo and 
Strogatz~\cite{Marvel-Mirollo-Strogatz-09}.

\subsection{WS variables versus order parameters}
For a hierarchically organized population we can define the mean field for each 
subpopulation; we call this quantity \textit{local mean field}.
If the subpopulation is finite, its mean field $Z_a$ is computed via 
Eq.~(\ref{orderpar}). 
For an infinitely large subpopulation it can be computed as
\begin{equation}
 Z_a=\int_0^{2\pi} w_a(\phi) e^{i\phi}d\phi\;,
\label{locordpar}
\end{equation}
where $w_a(\phi)$ is the probability distribution density of phases in the 
subpopulation $a$ (cf. Section~\ref{tl1}).

Let us relate $Z_a$  and WS variables. [For simplicity of presentation we omit the index 
$a$ in the following equations in this subsection.]
With the help of Eq.~(\ref{ws-trans}) we obtain:
\begin{equation}
 Z=r e^{i\Theta}=N^{-1}\sum_{k=1}^{N} e^{i\phi_k}=
\rho e^{i\Phi}\gamma(\rho,\Psi)=z\gamma(z,\alpha)\;,
\label{zKz}
\end{equation}
where
\begin{equation}
 \gamma(\rho,\Psi)=N^{-1}\sum_{k=1}^{N}\frac{1+\rho^{-1}e^{i(\psi_k-\Psi)}}
{1+\rho e^{i(\psi_k-\Psi)}}\;,
\label{gamma}
\end{equation}
or, using another set of variables,
\begin{equation}
\gamma(z,\alpha)=N^{-1}\sum_{k=1}^{N}\frac{1+|z|^{-2}z^{*}e^{i(\psi_k+\alpha)}}
{1+z^{*}e^{i(\psi_k+\alpha)}}\;.
\label{gamma1}
\end{equation}
If the subpopulation is infinite, then  the summation is 
substituted by the integration and we have, e.g., instead of \reff{gamma} 
\begin{equation}
 \gamma(\rho,\Psi)=\int_{-\pi}^{\pi}\frac{1+\rho^{-1}e^{i(\psi-\Psi)}}
{1+\rho e^{i(\psi-\Psi)}}\sigma(\psi) d\psi\;.
\label{gammaint}
\end{equation}
We see that in general a relation between the WS variables and the order
parameter is rather complex and contains not only the macroscopic
variables $\rho,\Psi$ but also all microscopic constants $\psi_k$ (or, for an infinite subpopulation, depends
heavily on the distribution $\sigma(\psi)$).

Now we discuss an important particular case when the order parameter
$Z$
can be expressed only through two WS variables $\rho$ and $\Phi$. 
For this purpose we re-write the function $\gamma$ (see Eq.~\ref{gamma}) 
as a series:
\begin{align*}
\gamma(\rho,\Psi)=&N^{-1}\sum_{k=1}^{N}\left [ 
\left (1+\rho^{-1}e^{i(\psi_k-\Psi)}\right )\sum_{l=0}^{\infty}
\left(-\rho e^{i(\psi_k-\Psi)}\right)^l
\right ] \\[1ex]
=&\left (\sum_{l=0}^{\infty}-\rho^l e^{-il\Psi}
\right )\left ( N^{-1}\sum_{k=1}^{N}e^{il\psi_k}\right)+\\[1ex]
&\rho^{-1}\sum_{l=0}^{\infty}(-\rho)^l e^{-i(l+1)\Psi}
\left (N^{-1}\sum_{k=1}^{N}e^{i(l+1)\psi_k} \right )\\[1ex]
=&\sum_{l=0}^{\infty} C_l \left (-\rho e^{i\Psi}\right )^l-
\rho^{-2}\sum_{l=0}^{\infty} C_{l+1} \left (-\rho e^{i\Psi}\right )^{l+1}\;,
\end{align*}
where the coefficients 
\begin{equation}
 C_l=N^{-1}\sum_{k=1}^{N}e^{il\psi_k}
\label{discr} 
\end{equation}
are the amplitudes of the  Fourier harmonics of the distribution of 
constants of motion $\psi_k$. 
Using that $C_1=0$ due to Eq.~(\ref{psicond}) or Eq.~(\ref{psicond1}), 
respectively, we finally obtain
\begin{equation}
 \gamma=1+(1-\rho^{-2})\sum_{l=2}^{\infty}C_l(-\rho e^{-i \Psi})^l \;.
\label{gamma_ser}
\end{equation}
The same series representation for $\gamma$ is obtained in the
thermodynamic limit when 
$\gamma$ is computed via integration (see Eq.~(\ref{gammaint})).
In this case the 
coefficients are computed according to
\[
 C_l=\int_{-\pi}^{\pi}\sigma(\psi)e^{il\psi}d\psi\;.
\]

The crucial observation is that Eq.~(\ref{gamma_ser}) essentially simplifies and we 
obtain simply $\gamma=1$ if all the amplitudes of the  Fourier harmonics
vanish, i.e. if $C_l=0$.
For an infinitely large system this is obviously true for  
a \textit{uniform distribution of constants of motion}, 
i.e. for $\sigma(\w)={2\pi}^{-1}$.
For a finite system this is valid approximately, if $N$ is large. 
Indeed, in this case for a uniform distribution of constants $\psi_k=\psi_1+2\pi(k-1)/N$ the
computation of $C_l$ according to \reff{discr} yields
$|C_l|=1$, $\mbox{arg}(C_l)=\psi_1 l$ for $l=N,2N,\ldots$, and $C_l=0$ otherwise, 
and we get 
\begin{equation}
\gamma=1+(1-\rho^{-2})\frac{\left [-\rho e^{i(\psi_1-\Psi)}\right ]^{N}}
{1-\left [-\rho e^{i(\psi_1-\Psi)}\right ]^{N}}\;.
\label{eq175}
\end{equation}
Hence, the deviation of $\gamma$ from unity decreases with the size of 
the subpopulation and, therefore, can be neglected for large $N$.

Using again the subpopulation index $a$, we summarize that for the uniform distribution 
of constants of motion and for large $N_a$
the order parameter of a subpopulation is directly expressed via the WS variables:
\begin{equation}
Z_a=\rho_a e^{i\Phi_a}=z_a \;. 
\label{op}
\end{equation}
As a result, the first two WS equations (see, e.g., Eqs.~(\ref{wsnce})) become equations 
for the amplitude and phase of the local mean field. The system simplifies further if
the forcing $H$ is independent of $\Psi$, then the third WS equation becomes irrelevant.

\subsection{WS variables versus generalized order parameters}
\label{secgamma1}


The Kuramoto order parameter $Z$ defined according to Eq.~\reff{orderpar} or
Eq.~\reff{locordpar} is an important quantity but it does not provide a full
characterization of the oscillator population. Such a characterization
is given by a set of
\textit{generalized Daido order parameters} \cite{Daido-92a,Daido-93a,Daido-95,Daido-96}. 
A generalized Daido
parameter of the order $m$ is defined according to
\be
Z_m=N^{-1}\sum_{k=1}^N e^{im\phi_k}\qquad \text{or} \qquad Z_m= \int_0^{2\pi}
w(\phi) e^{im\phi}d\phi\;.
\ee{genop}
Clearly, $Z_1$ is just the Kuramoto order parameter $Z$. The physical meaning
of the parameters $Z_m$ is especially transparent in the thermodynamic
limit: they are just the Fourier harmonics of the distribution of the
phases and thus completely characterize this distribution. 
Using \reff{ws-trans} we obtain:
\begin{equation}
 Z_m=\rho^m e^{im\Phi}\gamma_m(\rho,\Psi)
=z^m\gamma_m(z,\alpha)\;,
\label{zKz1}
\end{equation}
where
\begin{equation}
 \gamma_m(\rho,\Psi)=N^{-1}\sum_{k=1}^{N}\left(\frac{1+\rho^{-1}e^{i(\psi_k-\Psi)}}
{1+\rho e^{i(\psi_k-\Psi)}}\right)^m
\label{gamma-1}
\end{equation}
or 
\begin{equation}
 \gamma_m(\rho,\Psi)=\int_{-\pi}^{\pi}\left(\frac{1+\rho^{-1}e^{i(\psi-\Psi)}}
{1+\rho e^{i(\psi-\Psi)}}\right)^m\sigma(\psi) d\psi\;;
\end{equation}
[for brevity we skip a similar expression for $\gamma_m(z,\alpha)$.]
 
It can be seen that for uniform distribution of constants of motion in
the thermodynamic limit, 
i.e. for $\sigma=(2\pi)^{-1}$, 
$\gamma_m=1$ for all $m$. To show this, we again write $\gamma$ as a series and obtain
\[
\gamma_m= \int_{-\pi}^{\pi}\left(1+\rho^{-1}e^{i(\psi-\Psi)}\right)^m 
\left [1+ \sum_{l=1}^{\infty}\left( -\rho e^{i(\psi-\Psi)}\right )^l\right ]^m \sigma d\psi\;.
\]
Computing the powers and performing multiplication we obtain a series where the first term
is $(2\pi)^{-1}\int_{-\pi}^{\pi}d\psi=1$ and all other terms are products of some 
functions of $\rho$ and $\Psi$ with the integrals 
$(2\pi)^{-1}\int_{-\pi}^{\pi}e^{iL\psi} d\psi=0$, where $L$ is an integer.  
Thus, for the special case of uniformly distributed constants of motion we obtain 
$\gamma_m=1$  and
\begin{equation}
 Z_m=z^m=Z^m\;.
 \label{gord}
\end{equation}

Returning back to the notations for subpopulations of oscillators, we
can write a general relation between the local order parameters and the
WS variables as
\begin{equation}
 Z_{a,m}=\rho_a^m e^{im\Phi_a}\gamma_{a,m}(\rho_a,\Psi_a)
=z_a^m\gamma_{a,m}(z_a,\alpha_a)\;.
\label{zKz2}
\end{equation}
For a particular case of uniformly distributed constants of motion this relation 
simplifies to 
\begin{equation}
 Z_{a,m}=z_a^m=Z_a^m\;.
 \label{gord1}
\end{equation}
We emphasize that all results of this Section are valid for the case of a population 
with a continuous distribution of parameters as well. In this case we deal with the order 
parameters 
\begin{equation}
  Z_m(x,t)=\int_0^{2\pi}w(x,\phi) e^{im\phi}d\phi
\label{genordpar}
\end{equation}
and functions $\gamma(x,\rho,\Phi)$. For uniformly distributed constants $\psi$
we have $\gamma_m(x)=1$ and $Z_m(x)=Z^m(x)$. 

\subsection{The Ott-Antonsen ansatz}

Ott and Antonsen \cite{Ott-Antonsen-08} treated the Kuramoto problem in the thermodynamic limit 
$N\to\infty$ with the help of the continuity equation \reff{conteq}.
Writing the density function $w(\w,\phi,t)$ as a Fourier series 
\[
 w(\w,\phi,t)=\frac{n(\w)}{2\pi}\left\lbrace 1+\left[ 
\sum_{m=1}^{\infty}f_m(\w,t)e^{-im\phi}+\mbox{c.c.}
\right] \right\rbrace \;,
\]
where c.c. denotes complex conjugate, Ott and Antonsen noticed that the 
continuity equation is fulfilled if the Fourier coefficients can be expressed as 
\be
 f_m(\w,t)=\left [F(\w,t)\right ]^m\;,
\ee{oaans}
where $F(\w,t)$ is some function.

Comparing this approach with the definition of generalized order
parameters \reff{genop}, we see that the quantities $f_m$ are exactly
these order parameters and the ansatz \reff{oaans} means that
\[
Z_m(\w,t)=[Z(\w,t)]^m
\]
Thus, the found particular class of solutions, which we denote as the OA 
reduced manifold,
\textit{exactly corresponds} to the special case where the generalized 
order parameters are expressed via the powers of the WS variable $z$
(see Eqs.~(\ref{gord},\ref{gord1}). This holds if the
distribution of the WS constants $\psi$ is uniform.

The OA ansatz can be alternatively presented as follows.
Let us consider a generalized order parameter $Z_m(\w,t)$ of a subpopulation with the 
frequency $\w$, see Eq.~(\ref{genordpar}), and compute its time derivative 
\[
 \dot Z_m=\int_0^{2\pi}\fracpd{w(\w,\phi,t)}{t}e^{im\phi}d\phi=
 im\int_0^{2\pi}w(\w,\phi,t)\dot\phi e^{im\phi}d\phi\; ;
\]
here we also used Eq.~(\ref{conteq}).
Substituting $ \dot\phi=\w+(He^{-i\phi}-H^*e^{i\phi})/2i$ we obtain
(cf. \cite{Pazo-Montbrio-09}):
\[
 \dot Z_m =i\w m Z_m +\frac{m}{2}(H Z_{m-1}-H^*Z_{m+1})\;.
\]
This (infinitely dimensional) system of ODE obviously simplifies 
if $Z_m=Z^m$; this solution exactly corresponds to the OA manifold. 
On the other hand, it corresponds to the particular solution of the 
WS equations for the uniform distribution of constants of motion $\gamma(\w)=1$.
In this case the WS equations and OA approach yield the same equation for the time 
evolution of the order parameter:
\[
 \dot Z(\w,t) =i\w Z+\frac{1}{2}(H -H^*Z^2)\;.
\]
This important issue is illustrated in the next Sections.

\subsection{Reduced solution of a system with continuous frequency distribution}

In the very recent publication \cite{Ott-Antonsen-09}, Ott and Antonsen have demonstrated that 
the reduced manifold \reff{oaans} is the only attractive 
one if the globally coupled system has a continuous distribution of
parameters. 
Following their ideas, we demonstrate how this property follows from our 
theory for a particular case of harmonic force $H$ with frequency $\nu$;
for this purpose we use the results of Section~\ref{dws}. 
We remind, that in the terms of WS approach, the 
reduced OA manifold corresponds to the case $\gamma(x)=1$. 

Suppose we are interested only in a global characterization of the dynamics, e.g.,
we compute \textit{the global mean field}, which is obtained via a summation or 
an integration over the whole population:
\begin{equation}
Y=\int\,dx\, n(x) Z(x)=\int\,dx\, n(x) \gamma(x)\rho(x)e^{i\Phi(x)}\;.
\label{gop-1} 
\end{equation}
Substituting here
\[
  \gamma(x)=1+(1-\rho(x)^{-2})\sum_{l=2}^{\infty}C_l(x)(-\rho(x) e^{-i
  \Psi(x)})^l \;,
\]
to be compared with Eq.~\reff{gamma_ser}, we obtain
\begin{align}
 Y &= \int dx n(x)\gamma(x) \rho(x) e^{i\Phi(x)}=\nonumber\\ 
&=
\int dx\; n(x) \rho e^{i\Phi}+
\sum_{l\geq 2} (-1)^l \int dx\; n(x) \rho e^{i\Phi}  
C_l (\rho^l-\rho^{l-2}) e^{-il\Psi}\label{gop-22} \;.
\end{align}

We argue, that the sum in Eq.~\reff{gop-22} 
eventually vanishes and only the first term remains important. 
In Section~\ref{dws} we have shown that for the case of a harmonic
forcing, two types of solutions 
are possible for a subpopulation with the frequency $\w(x)$: 
a fully synchronous attractive solution with $\rho=1$ and a quasiperiodic solution
described in Appendix~\ref{AppWE}. 
For the  interval(s) of parameter $x$ where $\rho=1$, all terms in the sum in
Eq.~\reff{gop-22} vanish. 
Thus, we have to consider only integrals over the intervals where
$\rho(x,t)<1$.
To this end it is important to note that the expression under the integrals in the sum 
contains oscillating functions $\rho$, $\Phi$, and $\Psi$. 
According to \reff{apWE-6}, the phase variables $\Phi,\Psi$ rotate with a frequency that 
smoothly depends on $x$. Hence, the expression under the integrals 
 oscillates with 
some frequency, smoothly depending on $x$.
Thus, for large $t$ this term rapidly oscillates in $x$ 
 and, hence, the integral over $x$ vanishes, provided the other
 dependencies on parameter $x$ are sufficiently smooth. 
Mostly important, the distribution of oscillators $n(x)$ should not have
singularities. Otherwise, if $n(x)$ contains $\delta$-functions, 
the integrals in the sum do not vanish.

Summarizing, we expect that for $t\to\infty$ the sum in Eq.~(\ref{gop-22}) tends to 
zero and the global order 
parameter can be expressed via the WS variables $\rho,\Phi$ only:
\be
 Y =\int d x\; n(x) \rho(x) e^{i\Phi(x)}\;.
\ee{gop-3}
Thus, the functions $\gamma(x)$ become irrelevant for the macroscopic dynamics. 
For $t\to\infty$ the mean field can be computed  as if $\gamma=1$, for all frequencies, 
what corresponds to the OA reduced manifold.
Physically, this occurs due to the effective ``collisionless'' mixing of different 
subpopulations that are not 
locked by the common force. This effect is similar to the Landau damping
in plasmas and to the inhomogeneous line broadening in optics. 

We note that our argumentation treats only the case of harmonic mean field, because 
it is based on solutions of the WS equations (\ref{ws1-1}-\ref{ws1-3}). 
A rigorous proof for arbitrary functions $\w$ and $H$ was given recently 
by  Ott and Antonsen~\cite{Ott-Antonsen-09}.

\section{Mean field coupling}
\label{sec:lin}

\subsection{Organization of a subpopulation: mean field coupling}

Till now we considered a general time dependent force $H_a$, acting on elements 
of the subpopulation $a$, just it had to act equally on all elements.
Now we specify this force and consider several popular models as particular examples 
of the general approach.

Generally, the force $H_a$ can have two components. 
The first one arises from the interaction between elements of the subpopulation 
$a$ itself (Fig.~\ref{popstructure}). Very often it is assumed that this component 
is computed in a mean field fashion, i.e. that the coupling within each pair of 
oscillators is the same. 
Hence, the component of the forcing due to internal interaction is proportional to the
 complex mean field (order parameter) $Z_a$ of the subpopulation 
(see Eq.~(\ref{orderpar})). The proportionality factor we denote by $E_{aa}$; generally it is 
complex.

The second component results from all forces external with respect 
to the subpopulation $a$: those are the forces from all other subpopulations, 
regular or noisy forces acting on the whole population, etc. 
Next usual assumption is that all elements of $a$ equally contribute to the force 
acting on other subpopulations, and, vice versa, and therefore the effect of subpopulation 
$b$ on subpopulation $a$ is proportional to its complex mean field $Z_b$ and 
to its size $n_b$, so that 
\begin{equation}
H_a=F_{a,ext} +\sum_{b=1}^M E_{ab}n_b Z_b\;.
\label{osub2}
\end{equation}
Here $F_{a,ext}$ is the sum of all other external forces, acting on $a$, 
complex constants $E_{ab}$ describe the coupling from subpopulation $b$ to subpopulation $a$ 
(the term in the sum with $a$=$b$ corresponds to the first component described above),
and $n_b=N_b/N$ are relative population sizes.\footnote{
For the following it is convenient not to absorb $n_b$ into the coupling constant $E$, 
but to keep the effect of the subpopulation size explicitely.
}
At this point we emphasize that coupling determined by Eqs.~(\ref{osub2}) is only a simplest 
form of the mean field coupling which we denote as \textbf{linear}. The general, \textbf{nonlinear}
mean field coupling will be discussed below  in Section~\ref{sec:nonlin}, 
and now we consider several examples of linearly coupled ensembles.

First of all we derive the closed set of WS equations for interacting 
subpopulations with the mean field coupling.
For the case of discrete set of subpopulations we complement 
Eqs.~(\ref{hp-1},\ref{hp-2}) by the expression for the force 
Eq.~(\ref{osub2}) which we re-write, using  Eq.~(\ref{zKz}), as 
\begin{equation}
 H_a=F_{a,ext} +\sum_{b=1}^M E_{ab}n_b \gamma_b z_b\;.
\label{gop-4} 
\end{equation}
The continuous analog of this equation is 
\begin{equation}
 H(x)=F_{ext}(x)+\int\,dy\,E(x,y)n(y) \gamma(y)z(y)\;.
 \label{gop-5}
\end{equation}
It complements Eqs.~(\ref{ws-c1},\ref{ws-c2}) in case of infinite 
number of subpopulations. In these relations $\gamma$ is defined
according to Eq.~(\ref{gamma}) or Eq.~(\ref{gammaint}). 
We stress here that the systems 
(\ref{hp-1},\ref{hp-2},\ref{gop-4}) and (\ref{ws-c1},\ref{ws-c2},\ref{gop-5})
represent an {\bf exact} reduction of the dynamical description of 
hierarchical populations of oscillators by virtue of the WS ansatz.

As already discussed, the obtained equations significantly simplify on the
Ott-Antonsen manifold.  
In Section~\ref{secgamma1} we have demonstrated that for the uniform distribution of 
the constants of motion of a subpopulation, the corresponding function $\gamma=1$. 
In this case one of the WS variables 
decouples and we obtain a reduced system 
 \begin{align}
\frac{dz_a}{dt}&=i\w_a z_a+\frac{1}{2}H_a-\frac{z_a^2}{2}H_a^*\;, \label{wsred-1}\\[1ex]
H_a&=F_{a,ext} +\sum_{b=1}^M E_{ab}n_b z_b\;,\label{wsred-2}
\end{align}
for a discrete set of interacting subpopulations, or, respectively,
\begin{align}
\frac{\partial z(x,t)}{\partial t}&=i\w(x,t) z(x,t)+\frac{1}{2}H(x,t)-\frac{z^2}{2}H^*(x,t)\;, 
\label{wsred-3}\\[1ex]
H(x,t)&=F_{ext}(x,t)+\int\,dy\,E(x,y)n(y) z(y,t)\;,
\label{wsred-4}
\end{align}
for a continuous distribution of parameters.

\subsection{Example 1: The Kuramoto-Sakaguchi model}
\label{sec:ksm}
Suppose that external forces are absent, i.e. in Eq.~(\ref{osub2}) $F_{a,ext}=0$ for all $a$.
Writing the coupling constants as $E_{ab}=\e_{ab}e^{i\beta_{ab}}$ and substituting  
Eqs.~(\ref{orderpar},\ref{osub2}) into Eq.~(\ref{hopo-3}) we come to the model
(cf. \cite{Barreto-Hunt-Ott-So-08,Ott-Antonsen-08}):
\begin{equation}
 \fracd{\phi_k^{(a)}}{t}=\w_a+\frac{1}{N}\sum_{b=1}^M\sum_{j=1}^{N_b}\e_{ab}
\sin(\phi_j^{(b)}-\phi_k^{(a)}+\beta_{ab})\;,\qquad
k=1,\ldots,N_a\;.
\label{hopo-7}
\end{equation}
If we furthermore assume that the coupling parameters are the same for all subpopulations,
$\e_{ab}=\e$, $\beta_{ab}=\beta$, then we obtain the well-known 
Kuramoto-Sakaguchi model~\cite{Kuramoto-84,Sakaguchi-Kuramoto-86} of 
globally coupled oscillators:
\begin{equation}
 \fracd{\phi_k^{(a)}}{t}=\w_a+\e\frac{1}{N}\sum_{l=1}^{N}\sin(\phi_l-\phi_k^{(a)}+\beta)\;,\qquad
k=1,\ldots,N_a\;,
\label{hopo-8}
\end{equation}
where the summation is over the whole population containing $N$ oscillators. 
In terms of the global mean field $Y=re^{i\Theta}$ the model reads
\begin{equation}
 \fracd{\phi_k^{(a)}}{t}=\w_a+\e r\sin(\Theta-\phi_k^{(a)}+\beta)\;,\qquad
k=1,\ldots,N_a\;,
\label{kursak}
\end{equation}
The analysis of this model by virtue of WS reduction for the case of identical 
oscillators has been performed in the original WS paper~\cite{Watanabe-Strogatz-94}.

We proceed here by discussing the Kuramoto-Sakaguchi model with a continuous 
frequency distribution; its description is given by Eqs.~(\ref{ws-c1},\ref{ws-c2},\ref{gop-5}).
For this model it is natural to identify the continuous variable $x$ with 
frequency $\w$.
For the common effective force we obtain
\begin{equation}
 H=EY=\e e^{i\beta}Y\;,
\label{kscomfor}
\end{equation}
where $Y$ is the global mean field (see Eq.~(\ref{gop-1})).
Substituting the force into Eqs.~(\ref{ws-c1},\ref{ws-c2}) we obtain a closed system of equations
\begin{align}
\frac{\partial z(\w,t)}{\partial t}&=i\w z+\frac{E}{2}Y-\frac{E^*}{2}z^2Y^*\;, 
\label{ws-c3}\\[1ex]
\frac{\partial\alpha(\w,t)}{\partial t}&=\w+\text{Im}\left ( z^*EY\right )\;.
\label{ws-c4}
\end{align}

Consider now the reduced set of equations for the Kuramoto-Sakaguchi problem. 
This reduced set corresponds to the case $\gamma(\w)=1$, $z(\w)=Z(\w)$ and 
contains only two equations
(\ref{ws-c3},\ref{kscomfor}). 
Next, we consider the Lorentzian distribution of natural frequencies, 
$n(\w)=[\pi(\w^2+1)]^{-1}$.
As demonstrated by Ott and Antonsen \cite{Ott-Antonsen-08}, 
for this case, under an additional assumption that $z(\w)$ is analytic
in the upper half-plane, the integral 
in Eq.~(\ref{gop-1}) can be calculated by the residue of the pole at $\w=i$; 
this calculation yields $Y=z(i)$. 
Substituting this along with $\w=i$ into Eq.~(\ref{ws-c3})  we 
obtain the OA equation for the time evolution of the  Kuramoto mean field: 
\begin{equation}
 \fracd{Y}{t}=\left (-1+\frac{E}{2}\right ) Y -\frac{E}{2}Y^2 Y^*\;.
\label{nc-3}
\end{equation}
This closed equation for the order parameter was first derived and
solved in~\cite{Ott-Antonsen-08}; the solution is
\begin{equation}
r(t)=R\left\lbrace 1+ \left[ \left (\frac{R}{r_0}\right ) ^2-1\right] e^{(2-\e)t} \right\rbrace ^{-1/2}\;,
 \label{OAeqsol}
\end{equation}
where $R=\sqrt{1-2/\e}$ 
(notice a misprint in Eq.~(11) of \cite{Ott-Antonsen-08}).  
However, generally solutions of \reff{nc-3} do not 
coincide with the solutions of the full equation system (\ref{ws-c3},\ref{ws-c4}). 
We illustrate this important issue by the following numerical examples.
 
\begin{figure}[!t]
\centerline{\includegraphics[width=0.6\textwidth]{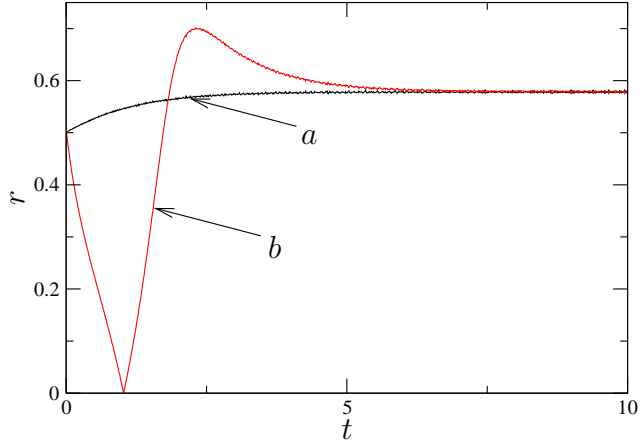}}
\caption{(Color online) Mean field amplitude $r$ as a function of time, 
for the Kuramoto ensemble with the Lorentzian distribution of frequencies; 
the number of oscillators is $N=10^4$, $\e=3$. 
Curves (a) and (b) correspond to two different sets of initial phases,
as described in the text. In the first case the evolution of the mean field
follows theoretical solution given by Eq.~\protect\reff{OAeqsol}, while for the second case
the transient dynamics deviates significantly from this solution.
}
\label{OAtest}
\end{figure}

First, we show that the solution deviates from \reff{OAeqsol} if the
analyticity assumtion above does not hold. We perform a direct 
numerical simulation of the Kuramoto-Sakaguchi model 
with $N=10^4$ oscillators. 
The frequencies of the oscillators are all different and are 
chosen to approximate the Lorentzian distribution. 
For the parameters of coupling we take $\beta=0$, so that $E=\e$ is real.
We perform two runs with the same  
\textit{macroscopic initial conditions} for the ensemble, 
choosing $Y(0)=r(0) e^{i\Theta(0)}=r_0=0.5$, but with different initial distribution 
of phases. 
Practically, we introduce an auxiliary angle variable $\varsigma$ which
attains $N$ values,
labelled by index $k$, 
uniformly distributed between $-\pi$ and $\pi$ (end points are excluded).
The frequencies of oscillators are then obtained as
$\w_k=\tan\ds\frac{\varsigma_k}{2}$
and the initial phases as
\begin{equation}
\phi_k=\pm 2\arctan\left[\frac{1-r_0}{1+r_0}\tan(\varsigma_k/2)\right ]=
\pm 2\arctan\left[\frac{1-r_0}{1+r_0}\w_k\right ]\;, 
\label{oatest1}
\end{equation}
cf. Eq.~(\ref{ws-trans1}); here the plus and minus signs correspond to the first and 
the second run, respectively.
Using Eq.~(\ref{ws-trans}) and Appendix~\ref{AppWS} we write for the WS variable
\begin{equation}
z(\w_k)=e^{i\phi(\w_k)}=\frac{r_0(1\mp\w_k)+1\pm i\w_k}{r_0(1\pm
i\w_k)+1\mp i\w_k}\;.
\end{equation}
(Notice that $\rho(\w_k)=1$ because we have only one oscillator at each frequency.)
Considering the obtained expression as an approximation of a continuous function 
$z(\w)$, we find that the latter has a pole at $\w=\mp i\frac{1+r_0}{1-r_0}$.
Thus, the first case corresponds to an initial condition that is
analytic in the upper half-plane, while the second run corresponds to
initial conditions that are analytic in lower half-plane.
The results are 
shown in Fig.~\ref{OAtest}; we see that the transient dynamics of the 
global mean field heavily depends on the microscopic initial conditions.
We emphasize that the result for the first set of initial conditions
very well agrees
with the solution \reff{OAeqsol},
while for the second set of initial 
conditions the transient dynamics is essentially different.

\begin{figure}[!t]
\centerline{\includegraphics[width=0.6\textwidth]{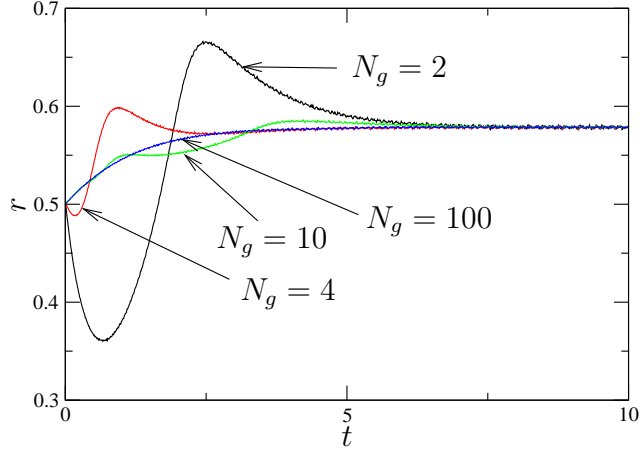}}
\caption{Mean field amplitude $r$ as a function of time, for $\e=3$, $M=10^4$. In all simulations
initial conditions have been chosen so that $r(0)=0.5$. For $N_g=100$ the evolution of the mean field
follows Eq.~\reff{OAeqsol}, while for smaller group sizes the transient deviates significantly 
from the OA manifold.
In all runs the distribution of the constants $\psi_{a,k}$ inside each group was chosen 
to be uniform ($q=1$).
}
\label{kurtrans1}
\end{figure}

Next, we verify the validity of the theory for hierarchically organized
populations, by simulating the same model 
with $M=10^4$ groups of identical elements. 
All groups have the same size, i.e. 
$N_a=N_g$ for all $a=1,\ldots,M$. Again, we always start with the same
macroscopic initial conditions for the ensemble, choosing $Y(0)=0.5$. 
However, the \textit{microscopic initial conditions}, given by the distribution of the 
constants of motion $\psi_{a,k}$, differ from run to run. In particular, we introduce a 
parameter $0<q\le 1$ that quantifies deviation of the distribution of $\psi_{a,k}$ from 
a uniform one; the value $q=1$ corresponds to the uniform distribution.
In Appendix~\ref{AppIC} we describe how one can choose different microscopic initial 
conditions while keeping the same macroscopic initial conditions. Here
we choose the initial state in such a way, that $z(\w)$ is analytic in
the upper half-plane.

\begin{figure}[!t]
\centerline{\includegraphics[width=0.6\textwidth]{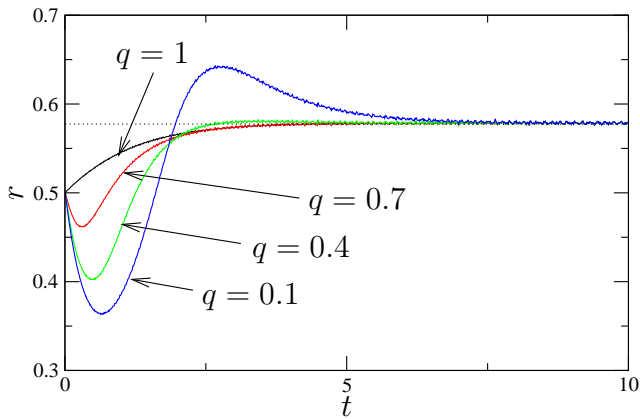}}
\caption{The same as Fig.~\ref{kurtrans1}, but 
for different distributions of constants of motion $\psi_{a,k}$ and fixed group size $N_g=200$.
Dotted line shows the theoretical asymptotic value of $r$.
}
\label{kurtrans2}
\end{figure}
First we analyze the effect of the subpopulation size $N_g$ (Fig.~\ref{kurtrans1}), 
taking $q=1$.
Theoretically, this effect is described with the help of Eq.~(\ref{eq175}). 
The results confirm the theoretical prediction: with increase of $N_g$ the transient
dynamics tends to the OA manifold and is nicely described by Eq.~\reff{OAeqsol},
while for small number of oscillator in a group,  
the deviations from the OA solution are essential. If there is one
oscillator in a group, the OA solution \reff{OAeqsol} is again valid
because here  $\rho=1$.
The effect of a non-homogeneous distribution of the microscopic constants $\psi_{k,a}$ 
on the dynamics of the mean field, for a large group sizes, 
is illustrated in Fig.~\ref{kurtrans2}. 
One can see that the deviations from the OA manifold become
larger as this distribution becomes less and less uniform, i.e. if the
parameter $q$ deviates from one. 
The examples of Fig.~\ref{OAtest},\ref{kurtrans1},\ref{kurtrans2} illustrate that although
the OA ansatz yields a simple closed system of equations, these equations do not describe 
the dynamics for general initial conditions, but only for a special subset of them.

\subsection{Example 2: Two coupled subpopulations}

For the next example we concentrate on a model, recently studied by Abrams, Mirollo,
Strogatz and Wiley \cite{Abrams-Mirollo-Strogatz-Wiley-08}.
They considered two identical subpopulations of the same size, 
i.e. $\w_1=\w_2=\w$ (without loss of generality we set it to zero) and
$N_1=N_2$, but the coupling within a subgroup differs from the coupling between the subgroups: 
$\e_{11}=\e_{22}=\mu$, $\e_{12}=\e_{21}=\nu\ne\mu$, and $\beta_{ab}=\beta$.
The equations are:
\begin{equation}
  \fracd{\phi_k^{(a)}}{t}=\w+
\mbox{Im}\left[\left(\mu Z_a+\nu Z_b\right)e^{i(\beta-\phi_k^{(a)})}\right ] \;,
\label{abrams}
\end{equation}
where $a=1,2$.
The WS system (\ref{hp-1},\ref{hp-2}) for this setup reads
\begin{align}
\frac{dz_1}{dt}&=\frac{1}{2}H_1-\frac{z_1^2}{2}H_1^*\;, \label{ab-1}\\[1ex]
\frac{d\alpha_1}{dt}&=\text{Im}(z_1^*H_1)\;,\label{ab-2}\\[1ex]
\frac{dz_2}{dt}&=\frac{1}{2}H_2-\frac{z_2^2}{2}H_2^*\;, \label{ab-3}\\[1ex]
\frac{d\alpha_2}{dt}&=\text{Im}(z_1^*H_2)\;,\label{ab-4}\\[1ex]
H_{1,2}&=(\mu Z_{1,2}+\nu Z_{2,1})e^{i\beta}\;,\label{ab-5}
\end{align}
and the relation between $Z_{1,2}$ and $z_{1,2}$ is given by Eq.~\reff{zKz}.
By applying the OA ansatz, i.e. by setting $Z_{1,2}= z_{1,2}$ we obtain a set of equations, 
originally derived in Ref.~\cite{Abrams-Mirollo-Strogatz-Wiley-08}.
Analyzing these equations, Abrams \textit{et al.} have obtained an interesting 
solution where one subpopulation is completely synchronized, $|z_1|=1$,
while the other one is only partially synchronized, $|z_2|<1$. 
Moreover, this partially synchronous state can be either steady, $z_2=const$, 
or time-periodic, i.e. $z_2$ is a periodic function of time.  
These regimes are called \textit{chimera} states.

The model of Abrams \textit{et al.} serves as a good illustration of the usefulness 
of the above described approach based on the \textit{exact} WS theory. 
A complete description of the dynamics for arbitrary initial conditions is given 
not by the OA equations, but by system (\ref{ab-1}-\ref{ab-5},\ref{zKz}). 
Correspondingly, the additional equations generally lead to an additional
time-periodicity for chimera states~\cite{Pikovsky-Rosenblum-08}: 
a steady-state solution becomes time-periodic (Fig.~\ref{chim1}), 
and a time-periodic state becomes quasiperiodic (Fig.~\ref{chim2}).
We notice, that in this case the solutions do not evolve towards the OA manifolds, 
because the  distribution of the oscillators' parameters is not continuous.

\begin{figure}[!t]
\centerline{\includegraphics[width=0.4\textwidth]{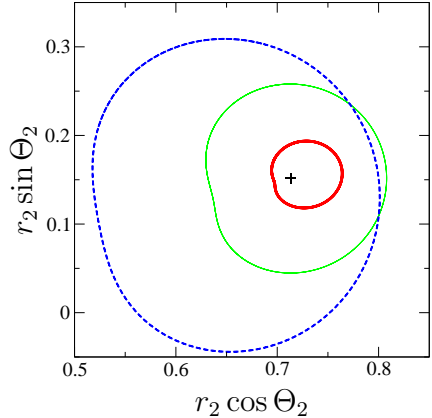}}
\caption{(Color online) Simulation of ensemble \reff{abrams} for $N=64$, 
$\beta=\pi/2-0.1$, $\mu=0.6$, 
$\nu=1-\mu=0.4$, and different distributions
of the microscopic constants $\psi_{k}^{(2)}$,
controlled by the parameter $q$ (see Appendix~\ref{AppIC}). 
Note that
the distribution of constants $\psi_{1,k}$ is irrelevant since
the first subpopulation is completely synchronized.
The case $q=1$ (marked by plus) corresponds to the OA manifold, 
here the mean field is constant.
For $q=0.9$, $q=0.7$, and $q=0.5$ one observes time-periodic states represented 
by limit cycles in the complex plane $Z_2$ 
(red bold, green solid,  and blue dotted curves, respectively). 
}
\label{chim1}
\end{figure}

\begin{figure}[!t]
\centerline{\includegraphics[width=0.6\textwidth]{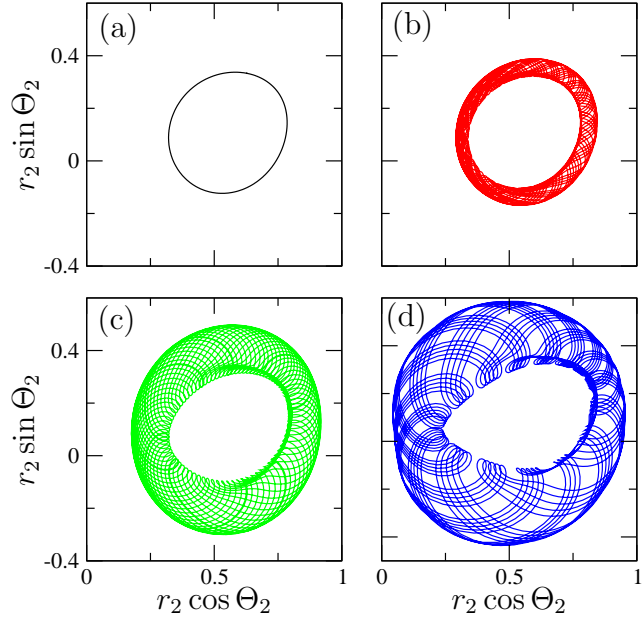}}
\caption{The same as in Fig.~\ref{chim1}, but for $\mu=0.65$ and $\nu=1-\mu=0.35$.
Now on the OA manifold, i.e. for  $q=1$,  the dynamics is periodic (a), 
while for more general initial conditions parameterized by 
 $q=0.9$ (b), $q=0.7$ (c), and  $q=0.5$ (d) the dynamics is quasiperiodic.
}
\label{chim2}
\end{figure}

\section{Nonlinear mean field coupling}
\label{sec:nonlin}

\subsection{General nonlinear coupling}
\label{sec:genlin}

Till this moment we restricted our consideration to the case when the force, acting on the 
oscillator subpopulation $a$, is a linear combination of the local mean fields $Z_b$ of 
all other subpopulations $b$ (see Eq.~(\ref{osub2})). 
We denoted this coupling as linear. 
Generally, this force can depend on the Daido generalized complex order 
parameters~\cite{Daido-92a,Daido-96}, see Eq.~(\ref{genop}).
For a general nonlinearity  we can expect that $H_a$ contains 
arbitrary combinations of $Z_{b,m}$. However, in a physically reasonable model some 
restrictions appear.

To discuss this important issue let us first consider an isolated subpopulation, say $a$, 
and recall that derivation of the phase models of type 
(\ref{hopo-1}-\ref{hopo-3}) incorporates an averaging over oscillation period 
(see, e.g., \cite{Kuramoto-84}). It means that the mean field forcing $H_a$ can 
include only the terms having the frequency $\approx \w$, i.e. the terms like
\[
 Z_{a,1}\;,\qquad Z_{a,m}Z_{a,1-m}\;,\qquad Z_{a,m}Z_{a,l}Z_{a,1-m-l}
\;,\ldots \;,
\]
where the sum of lower indices is one (we remind that $Z_{-m}=Z_m^*$). 
Let us now consider a particular case when the nonlinear mean field coupling is 
determined by the first order parameter $Z_a=Z_{a,1}=r_ae^{i\Theta_a}$ only. 
Remarkably, this case corresponds to OA reduced 
manifold since the latter implies $Z_m=Z^m$. The forcing then takes the form:
\begin{equation}
H_a=h_1 Z_a+h_3 |Z_a|^2 Z_a+h_5 |Z_a|^4 Z_a+\ldots \;.  
\label{hopo-11}
\end{equation}
Denoting $|h_1|=\e$ and $1+\frac{h_3}{|h_1|}|Z_a|^2+\frac{h_5}{|h_1|}|Z_a|^4+\ldots=
A(r_a,\e)e^{i\beta(r_a,\e)}$
we obtain 
\begin{equation}
 H_a=\e A(r_a,\e)e^{i(\beta(r_a,\e)}Z_a \;
\end{equation}
and
\begin{equation}
 \dot\phi_k^{(a)}=\w_a+\e A(r_a,\e)r\sin\left(\Theta_a-\phi_k+\beta(r_a,\e)\right)\;.
\end{equation}
This nonlinear model was suggested and analyzed in 
\cite{Rosenblum-Pikovsky-07,Pikovsky-Rosenblum-09}; 
its interesting dynamics is discussed below.

In case when the subpopulation $a$ interacts with the other populations and 
external fields,  Eq.~(\ref{hopo-11}) includes additional terms. The terms, describing 
interaction with other subpopulations can contain any combination of $Z_{b,m}$, with 
$b=1,\ldots,M$, however, only the terms having the frequency $\approx \w$ are resonant and 
therefore essential for the dynamics. 
This is the consequence of the fact that our basic model \reff{hopo-1}
is not general but contains only the first harmonics.

\subsection{A minimal model of nonlinearly coupled oscillators}

Our generalization \cite{Rosenblum-Pikovsky-07,Pikovsky-Rosenblum-09}
of the model (\ref{hopo-8}) accounts for a possible nonlinear 
response of the oscillator to the forcing.
It means that the effect of the large force is not just an ``up-scaled'' effect 
of the small one, but can be qualitatively different. 
For a more detailed explanation of this concept, 
let us consider one oscillator, influenced by a harmonic force with 
the amplitude $\delta$ and phase $\Theta$ and let the interaction be described by 
the sine-function, so that the equation for the oscillator's phase reads:
\begin{equation}
 \dot \phi=\w+A\delta\sin(\Theta-\phi_k+\beta)\;.
\label{forcedosc}
\end{equation}
Parameters $A$ and $\beta$ determine the response to the forcing. So, e.g., 
if $A\cos\beta>0$, the force with a frequency close to $\w$ results in a stable
in-phase synchronization of the oscillator. 
Consider now an ensemble of \textit{identical} oscillators, coupled via the mean field.
Comparing Eq.~(\ref{forcedosc}) with Eq.~(\ref{kursak}) 
we identify  $\delta$ with $\e r$ and $\Theta$ with the 
phase of the mean field.
If the oscillators are synchronized, the mean field has the same frequency 
and the phase as each of them, and if the above condition $A\cos\beta>0$ is fulfilled, 
then this synchronous regime is stable. 
Otherwise, if $A\cos\beta<0$,  the in-phase synchrony is unstable
and the oscillators remain asynchronous, i.e. they have different phases 
(the frequencies are the same since the oscillators are identical). 

Nonlinearity of the coupling means that the parameters in Eq.~(\ref{forcedosc}) 
can depend on the  amplitude of the force, 
$A=A(\delta)$, $\beta=\beta(\delta)$.\footnote{
Generally, functions $A$ and $\beta$ can depend not only on the product of $r$ 
and $\e$, but on both variables, i.e. $A=A(r,\e)$, $\beta=\beta(r,\e)$, see 
\cite{Pikovsky-Rosenblum-09} for details.}
In this case we can expect interesting dynamics to occur. Suppose the factor 
$A\cos\beta$ is positive for small $\delta$
but becomes negative when $\delta$ increases and achieves some critical 
value $\delta_c$. Then the synchronous state becomes unstable and the oscillators
tend to desynchronize. However, this immediately reduces the mean field, i.e. the 
amplitude of the forcing $\delta$, and the synchronous state becomes stable again, the 
oscillators tend to synchronize, what increases the mean field. As a result of
the counter-play of these two tendencies, the system settles at the border of 
stability, exhibiting partially synchronous dynamics: the (identical) 
oscillators are not synchronized, because synchrony is unstable, but they are 
also not completely asynchronous, because this state is unstable as well.
As a result, the oscillators distribute around the unit 
circle, forming a bunch.
Two different dynamical states appear depending on whether $A$ or $\cos\beta$ 
becomes negative with an increase of $\delta$.
In the former case this bunch remains static in the frame, rotating with the 
oscillator frequency $\w_{osc}$.
In the latter case, this bunch rotates with respect to 
the mean field, i.e. the oscillators have frequency that is generally incommensurate 
with the frequency of the mean field.
Furthermore, the bunch can also ``breath'' so that the mean field is modulated,
 see \cite{Rosenblum-Pikovsky-07,Pikovsky-Rosenblum-09}
for more details.

Let us now take an infinitely large population with a continuous frequency distribution.
Consider a particular case of the homogeneous coupling when all oscillators
are driven by the same force. Then we should omit the index $a$ 
in Eq.~(\ref{hopo-11}) and substitute there $Z_a$ by  the global mean field 
\begin{equation}
Y=r e^{i\Theta}=\int n(\w) Z(\w) d\w\;,\qquad  
 H=\e A(r,\e)e^{i\beta(r,\e)}Y \;
 \label{hy}
\end{equation}
This corresponds to the following microscopic equations
\begin{equation}
 \dot\phi_k=\w_k+\e A(r,\e)r\sin\left(\Theta-\phi_k+\beta(r,\e)\right)\;.
\label{soqmod}
\end{equation}
Next, we consider the infinitely large system and want to 
write the corresponding WS 
equations.
We emphasize that in the derivation of Eqs.~(\ref{ws-c3},\ref{ws-c4}) we did not 
assume that $E=\mbox{const}$; we only used the 
assumption that the coupling is homogeneous, i.e. that $E$ is independent of the 
frequency. Hence, we just have to substitute in Eqs.~(\ref{ws-c3},\ref{ws-c4}) 
$E$ with $\e A(r,\e)e^{i(\beta(r,\e)}$. This yields, together with
Eq.~\reff{hy}, a system of integral equations with nonlinear coupling. 
This full system is still rather difficult to analyse, therefore below we
perform, like in Section~\ref{sec:ksm} a further simplification 
allowing us to obtain an analog of Eq.~\reff{nc-3}.

\subsection{Nonlinearly coupled ensemble with the Lorentzian frequency distribution: Theory}
\label{nonlinlor}

In this Section we exploit the general theory to analyze in details the dynamics of 
ensembles with global nonlinear coupling (see Eqs.~(\ref{soqmod})) 
and Lorentzian distribution of frequencies.
Looking for the asymptotic solutions we follow the argumentation of 
Section~\ref{OAansatz} and consider the reduced dynamics, corresponding 
to the case $\gamma(\w)=1$ and, respectively, to the OA reduced manifold.
Furthermore, following \cite{Ott-Antonsen-08} we 
consider the Lorentzian distribution of natural frequencies, 
$n(\w)=[\pi(\w^2+1)]^{-1}$, and assume that the field $z(\w)$ is
analytic in the upper half-plane. Then, in a way similar to the
derivation of Eq.~\reff{nc-3}, we obtain
\begin{equation}
 \fracd{Y}{t}=\left (-1+\frac{\e A(r,\e)e^{i\beta(r,\e)}}{2}\right ) Y -
\frac{\e A(r,\e)e^{i\beta(r,\e)}}{2}Y^2 Y^*\;.
\label{nc-nl}
\end{equation}
Below we verify the validity of this equation by numerics; 
in particular we confirm that the asymptotic solutions are 
confined to the OA manifold. However, the basins of attraction of these solutions depend
on the distribution of the microscopic constants of motion.

Separating the real and imaginary parts of Eq.~(\ref{nc-nl}) we obtain equations for 
the amplitude $r$ and frequency $\W$ of the mean field:
\begin{align}
\fracd{r}{t}&= -r +\frac{\e A}{2}r(1-r^2) \cos\beta \;,  \label{nc-4} \\[1ex]
\fracd{\Theta}{t}&=\W=\frac{\e A}{2}(1+r^2) \sin\beta \;.\label{nc-5}
\end{align}
Stability of the asynchronous state $r=0$ of the ensemble is determined 
by the condition
\begin{equation}
\left . \frac{d\dot r}{dr}\right |_{r=0}=-1+\frac{\e}{2}\left . A\right|_{r=0}
\left . \cos\beta\right |_{r=0} 
=-1+\frac{\e}{2}A_0\cos\beta_0 <0\;,
\end{equation}
what yields the value of the critical coupling
\begin{equation}
 \e_{cr}=\frac{2}{A_0\cos\beta_0}\;.
\label{crcoupl}
\end{equation}
For $\e>\e_{cr}$ the ensemble exhibits a synchronous state with the mean field amplitude 
$0<r<1$. \footnote{It can be shown that for a physically reasonable model 
(see Section~\ref{sec:genlin}), the order parameter grows
 close to the transition point as
$r\sim\sqrt{\e-\e_{cr}}$, as in case of the standard Kuramoto model.
}
The latter is determined from the condition $\dot r=0$, which yields the equation
\begin{equation}
\e(1-r^2) A(r)\cos\beta(r)=2\;.
\label{eq_r}
\end{equation}
If the solution of this equation for particular functions $A$, $\beta$ is found 
(most likely, numerically), then Eq.~(\ref{nc-5}) provides the frequency of the 
mean field $\W$.
Note that the collective oscillation arises with the frequency 
$\W\left |_{\e=\e_{cr}}\right .=\tan(\beta_0)$. 
Below, we illustrate the theory by two particular choices of the amplitude $A(r,\e)$ and phase 
$\beta(r,\e)$ dependencies.

\subsection{Nonlinearly coupled ensemble with the Lorentzian frequency distribution: 
amplitude nonlinearity}
First we consider the impact of the amplitude function $A$, by setting
$A(r)=\mu-\e r^2$, $\beta=0$.
Equations (\ref{nc-5}) and (\ref{crcoupl}) yield $\W=0$ and $\e_{cr}=2/\mu$. 
From Eq.~(\ref{eq_r}) we obtain a quadratic equation for $r^2$
\begin{equation}
\e^2 r^4-\e(\e+\mu)r^2+\e\mu-2=f(r^2)=0\;.
\end{equation}
It is easy to see that for $\e>0$ we have $f(\pm\infty)=\infty$, $f(0)=\e\mu-2$, 
and $f(1)=-2$. 
Hence, for $\mu>0$, the solution in the $[0,1]$ interval exists for $\e>\e_{cr}=2/\mu$, 
it is given by 
\begin{equation}
r^2 =\frac{\e+\mu-\sqrt{(\e-\mu)^2 +8}}{2\e} \;.
\end{equation}
The dependence of $r$ on $\e$ for $\mu=2$, $\e_{cr}=1$ is shown by bold line 
in Fig.~\ref{bunfig1}.
\begin{figure}[!t]
\centerline{\includegraphics[width=0.6\textwidth]{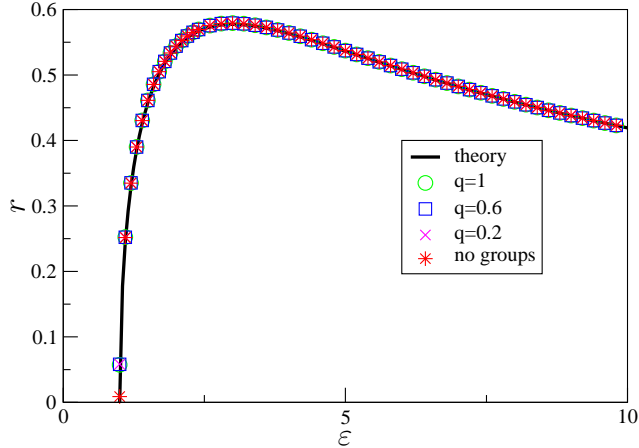}}
\caption{Mean field amplitude $r$ as a function of coupling strength $\e$ for 
the nonlinearly coupled ensemble with $A=2-\e r^2$ and $\beta=0$. 
Black bold line: theory.  Symbols show simulations of hierarchical ensembles with 
$M=1000$ groups of $20$ oscillators, for different initial conditions (see text).
Red line corresponds to simulation of a population, where all elements have 
different frequencies, without groups, and with a uniform distribution of $\phi$.
}
\label{bunfig1}
\end{figure}

We verify the theory by simulation of ensemble dynamics for $M=1000$ groups of $N_a=20$ 
oscillators. 
For this goal, we prepare different initial conditions, corresponding to uniform and 
nonuniform distribution of $\psi_{k,a}$; these distributions are parameterized by 
parameter $q$, so that $q=1$ and $q<1$ correspond to uniform and non-uniform 
distributions, respectively (see Appendix~\ref{AppIC}).
Next, we simulate the ensemble 
of $N=10000$ oscillators with different frequencies 
(in other words, each group contains only one oscillator); in this case we take for initial
conditions a nearly uniform distribution of phases $\phi_k$. We see, that the results, 
shown in Fig.~\ref{bunfig1}, demonstrate a good correspondence to the theory.

\subsection{Nonlinearly coupled ensemble with the Lorentzian frequency distribution: 
phase nonlinearity}
Now we analyze the effect of the dependence $\beta=\beta(r)$,
by setting $A=1$, $\beta=\beta_0+\e^2r^2$.
For the chosen particular function  we have 
$\e_{cr}=2/\cos\beta_0$ and Eq.~(\ref{eq_r}) yields an 
equation for $r^2$:
\begin{equation}
\e (1-r^2)\cos(\beta_0+\e^2 r^2)-2=f(r^2)=0 \;.
\label{eq_multistab}
\end{equation}
It is easy to see that $f(0)=(\e-\e_{cr})\cos\beta_0>0$ and $f(1)=-2$, 
hence there always exist at least one solution.
(We remind that $\beta_0=\mbox{const}$, $|\beta_0|<\pi/2$.)
Numerical analysis of Eq.~(\ref{eq_multistab}) shows that the number of its roots 
increases with $\e$. Thus, the system exhibits multistabilty.
The corresponding bifurcation diagram in the parameter plane $\beta_0,\e$ 
is shown in Fig.~\ref{bif_diag}.
\begin{figure}[ht!]
\centerline{\includegraphics[width=0.5\textwidth]{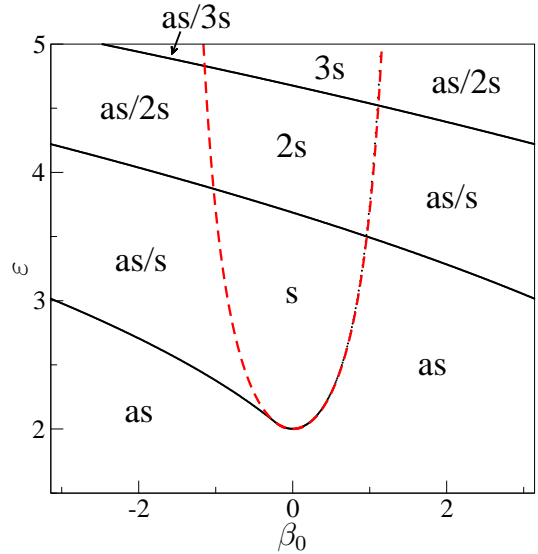}}
\caption{Multistability in the nonlinearly coupled ensemble with $A=1$, $\beta=\beta_0+\e^2r^2$.
Red dashed line shows critical coupling $\e_{cr}=2/\cos\beta_0$; inside the domain, determined 
by this curve, the asynchronous state is unstable. Labels \textbf{as}, \textbf{s}, 
and \textbf{ns} mean asynchrony 
(the state with $r=0$ is stable), synchrony ($r>0$), and coexistence of $n$ 
synchronous states, respectively. 
Label \textbf{as/ns} means coexistence of asynchronous and $n$  synchronous solutions. 
}
\label{bif_diag}
\end{figure}
Dependencies of the mean field amplitude  and frequency on the coupling strength 
for $\beta_0=0$  are shown in Fig.~\ref{r_om_beta0}.

\begin{figure}[ht!]
\centerline{\includegraphics[width=0.6\textwidth]{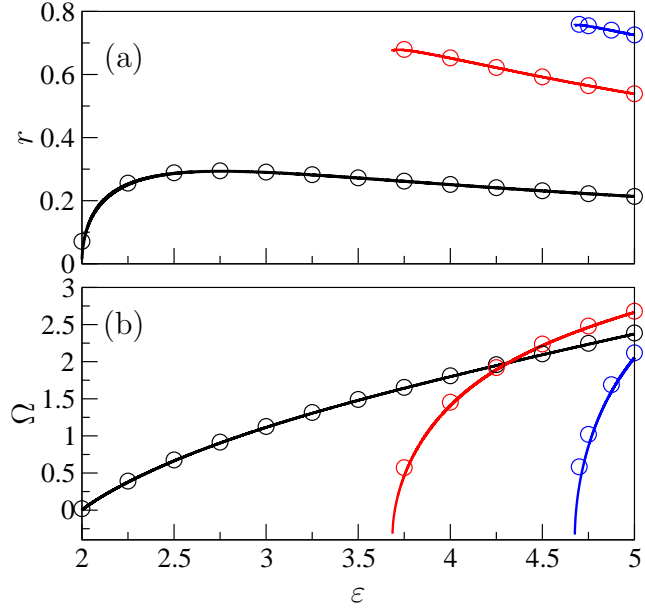}}
\caption{Illustration of the multistability in the nonlinearly coupled ensemble with $A=1$, 
$\beta=\beta_0+\e^2r^2$, for $\beta_0=0$. Three branches of the solution of 
Eq.~(\protect\ref{eq_multistab}) for the mean field amplitude are shown by different 
colors in (a). 
Corresponding solutions for the frequency of the mean field are 
shown by the same colors in (b). Numerical results (see text for details) are shown by 
symbols.
}
\label{r_om_beta0}
\end{figure}

To verify the theory we again perform a direct numerical simulation of an 
ensemble with $M=5000$ subpopulations of 
$N_a=20$ oscillators each, for $\beta_0=0$. 
The numerical results are shown by symbols in Fig.~\ref{r_om_beta0}.

Finally, we demonstrate that although the stationary dynamics of the system corresponds to 
uniform distribution of microscopic constants 
$\psi$ (i.e., to $q=1$), the transient dynamics 
does depend on the distribution. In other words, the attractors of the multistable 
system can be obtained by the simplified theory, see \cite{Ott-Antonsen-09} 
and discussion above, but their basins of attraction depend on the distributions 
of $\psi$. 
In numerical experiments, we simulate an ensemble containing $M=5000$ subpopulations of $20$
oscillator each, for $\e=4.5$,  taking $\rho_0=0.52$ and different values of $q$. 
The results shown in Fig.~\ref{mstab} demonstrate that starting from the same macroscopic 
initial conditions, the system can evolve to different attractors,
depending on the microscopic constants.

\begin{figure}[ht!]
\centerline{\includegraphics[width=0.6\textwidth]{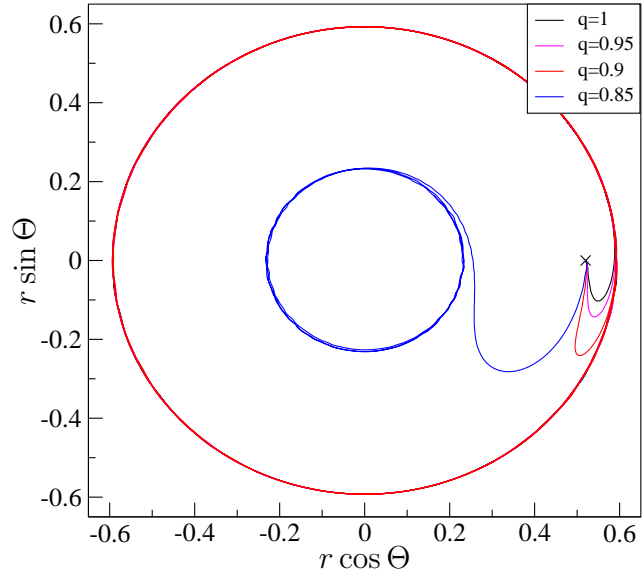}}
\caption{Evolution of the mean field for same macroscopic initial conditions 
($r=0.52$, $\Theta=0$, shown by a cross) but for different distributions of the constants of 
motions, parameterized by $q$ (see text). Note that both attractors of the system 
(limit cycles with the radius $\approx 0.23$ and $\approx 0.6$) correspond to the theory, 
developed under assumption of the uniform distribution of the constants of motion 
(cf. Fig.~\protect\ref{r_om_beta0}). However, transient and basins of attraction depend 
on the distribution of the constants of motion. 
}
\label{mstab}
\end{figure}

\section{Conclusions and outlook}
\label{conc}

The main goal of this paper was to provide a generalization of the powerful 
Watanabe-Strogatz theory on the heterogeneous populations of phase oscillators. 
We have formulated the Watanabe-Strogatz equations for a general hierarchically 
organized ensemble, and have examined limiting cases of infinite populations. 
Remarkably, there exist two possible thermodynamic limits:
in the first one we treat a finite number of infinitely large populations, whereas
in the second case we deal with a system with a continuous distribution of parameters, 
e.g., of frequencies.
The derived equations provide an exact reduction of the dynamics;
in many cases the problem under consideration becomes low-dimensional.
We have analyzed the derived equation in several important cases, including the 
Kuramoto-Sakaguchi model and the model of two coupled populations with chimera.
Noteworthy, the reduced equations are valid both for linear and nonlinear coupling; 
in the latter case the approach has allowed us 
to describe multistable synchronous dynamics. 

Next, we have thoroughly studied a relation between the Watanabe-Strogatz 
theory and the recent Ott-Antonsen ansatz and have demonstrated that the latter 
corresponds to a particular choice of initial conditions for the ensemble. To be exact,
the OA approach corresponds to the case when the constants of motion in 
the Watanabe-Strogatz ansatz are uniformly distributed. 
Although the Ott-Antonsen equations are much simpler than the full Watanabe-Strogatz 
system, several examples considered have shown that they provide only asymptotic
solutions, whereas the transient dynamics and the basins of attraction of these solutions
depend on the choice of initial conditions. 
(See \cite{Ott-Platig-Antonsen-Girvan-08} for another example of
 nontrivial transient dynamics off the OA manifold.)

Finally, we would like to mention that the approach presented opens 
new perspectives in analysis of such long-standing problems as
finite-size effects and the effects of a common external noise on 
oscillator ensembles. Also application of it to systems with delayed
coupling appears promising.

We thank A. Politi and E. Ott for useful discussions, and E. Ott and S. Strogatz 
for communicating their works prior to publication.  
The work was supported by DFG
(SFB 555).

\clearpage
\appendix
\section{Watanabe-Strogatz equations in new notations}
\label{AppWS}

According to Watanabe and Strogatz~\cite{Watanabe-Strogatz-93,Watanabe-Strogatz-94}, the system of $N>3$ identical globally coupled phase 
oscillators 
\be
\fracd{\phi_k}{t}=\w(t)+A(t)\sin(\xi(t)-\phi_k)=\w(t)+ g(t)\cos\phi_k+h(t)\sin\phi_k \;
\ee{ap1-1}
admits a low-dimensional description.
For arbitrary functions of time $\w(t)$, $g(t)$, and $h(t)$,
this $N$-dimensional system is completely described by three global 
variables plus constants of motion $\psi_k$, $k=1,\ldots,N$, 
which obey three additional constraints, so that $N-3$ of them are independent. 
The ``global phases'' $\tilde\Psi$ and $\tilde\Phi$ and 
the global ``amplitude'' $0\le\tilde\rho\le 1$ obey the WS equations
\begin{align}
\dot{\tilde\rho} &=-(1-\tilde\rho^2)(g\sin\tilde\Phi-h\cos\tilde\Phi)\label{eq4-1}\;,\\[1ex]
\tilde\rho\dot{\tilde\Psi}&=
-\sqrt{1-\tilde\rho^2}(g\cos\tilde\Phi+h\sin\tilde\Phi)\label{eq4-2}\;,\\[1ex]
\tilde\rho\dot{\tilde\Phi}&= -g\cos\tilde\Phi-h\sin\tilde\Phi \;.
\label{ap1-2}
\end{align}
The solution of the original system (\ref{ap1-1}) can be recovered via the 
following transformation:
\begin{equation}
 \tan\left ( \frac{\phi_k-\tilde\Phi}{2}\right ) =\sqrt{\frac{1+\tilde\rho}{1-\tilde\rho}}
\tan\left ( \frac{\psi_k-\tilde\Psi}{2}\right )\;.
\label{ap1-3}
\end{equation}

We perform the variable substitution $\tilde\rho,\tilde\Psi,\tilde\Phi\to \rho,\Psi,\Phi$ 
according to 
\begin{equation}
 \tilde\rho=\frac{2\rho}{1+\rho^2}, \quad \tilde\Psi=\Psi+\pi, \quad \tilde\Phi=\Phi+\pi\;.
\end{equation}
Rewriting the r.h.s. of Eq.~(\ref{ap1-1}) as $\w(t)+\mbox{Im}\left (Z(t) e^{-i\phi_k}\right )$ 
with obvious relations $\mbox{Re}(Z)=-h(t)$, $\mbox{Im}(Z)=g(t)$, we obtain the system of WS 
equations (\ref{ws-1}-\ref{ws-3}) in new variables.
The transformation (\ref{ap1-3}) now takes the form
\begin{equation}
 \tan\left ( \frac{\phi_k-\Phi}{2}\right ) =\frac{1-\rho}{1+\rho}
\tan\left ( \frac{\psi_k-\Psi}{2}\right )\;.
\label{ap1-7}
\end{equation}
It is convenient to re-write this transformation in the exponential form, using the 
following identity:
\begin{align*}
 e^{i\alpha}&=\frac{1+i\tan(\alpha/2)}{1-i\tan(\alpha/2)}=
\frac{\left [1+i\tan(\alpha/2)\right ]^2}{1+\tan^2(\alpha/2)}\\[1ex]
&=\cos^2\frac{\alpha}{2}\cdot\left ( 1-\tan^2\frac{\alpha}{2}+2i\tan\frac{\alpha}{2}\right )
=\cos\alpha+i\sin\alpha\;.
\end{align*}
With the help of this identity we write:
\begin{align*}
 e^{i(\phi_k-\Phi)}&=\frac{1+i\tan\frac{\phi_k-\Phi}{2}}{1-i\tan\frac{\phi_k-\Phi}{2}}
=\frac{1+i\frac{1-\rho}{1+\rho}\tan\frac{\psi_k-\Psi}{2}}
{1-i\frac{1-\rho}{1+\rho}\tan\frac{\psi_k-\Psi}{2}}\\[1ex]
&=\frac{(1+\rho)\cos\frac{\psi_k-\Psi}{2}+i(1-\rho)\sin\frac{\psi_k-\Psi}{2}}
{(1+\rho)\cos\frac{\psi_k-\Psi}{2}+i(\rho-1)\sin\frac{\psi_k-\Psi}{2}}\\[1ex]
&=\frac{\rho e^{-i(\psi_k-\Psi)/2}+e^{i(\psi_k-\Psi)/2}}
{\rho e^{i(\psi_k-\Psi)/2}+e^{-i(\psi_k-\Psi)/2}}\;,
\end{align*}
what yields the desired transformation (\ref{ws-trans}).

\section{Dynamics of WS variables in an external field}
\label{AppWE}

Here we present the exact solution of Eqs.~(\ref{ws2-1}-\ref{ws2-3}) for the case
$|\w-\nu|>H_0$. The system has one steady state $\Delta_0=0$ and
\begin{equation}
\rho_0=\begin{cases}
-(\w-\nu)H_0^{-1}+\sqrt{(\w-\nu)^2H_0^{-2}-1}&\mbox{for~}(\w-\nu)>H_0\;,\\[1ex]
-(\w-\nu)H_0^{-1}-\sqrt{(\w-\nu)^2H_0^{-2}-1}&\mbox{for~}(\w-\nu)<-H_0\;.
\end{cases}
\label{fixp}
\end{equation}
Equation~\reff{ws2-3} then yields
\[
 \dot\alpha=\w-\nu-H_0\rho_0=2(\w-\nu)\mp\sqrt{(\w-\nu)^2-H_0^{2}}\;.
\]
Hence, $\Psi$ rotates with the frequency 
\begin{equation}
 \kappa=3\nu-2\w\pm\sqrt{(\w-\nu)^2-H_0^{2}}\;.
\label{fixpfreq}
\end{equation}

Consider now solution with $\rho\ne\rho_0$ and $\Delta\ne\Delta_0$.
Introducing new variables $u=\rho\cos\Delta$, $v=\rho\sin\Delta$, we write 
the first two equations as
\begin{equation}
\begin{aligned}
\dot u&=-(\w-\nu)u-H_0uv\;,\\
\dot v&=(\w-\nu)u+\frac{H_0}{2}(1+u^2-v^2)\;,
\end{aligned}
\label{apWE-1}
\end{equation}
with $v_0=0$ and $u_0=\rho_0$.
Introducing $w=u-u_0$ we rewrite \reff{apWE-1} as
\begin{equation}
\begin{aligned}
\dot w&=-(\w-\nu+H_0u_0)v-H_0wv\;,\\
\dot v&=(\w-\nu+H_0u_0)w+\frac{H_0}{2}(w^2-v^2)\;.
\end{aligned}
\label{apWE-2}
\end{equation}
Using an ansatz $p=v(v^2+w^2)^{-1}$, $q=w(v^2+w^2)^{-1}$ and denoting
\begin{equation}
\kappa=\w-\nu+H_0u_0= \mp\sqrt{(\w-\nu)^2-H_0^{2}}
\end{equation}
we simplify the equations to
\begin{equation}
\begin{aligned}
\dot p&=\kappa q+\frac{H_0}{2}\;,\\
\dot q&=-\kappa p\;.
\end{aligned}
\label{apWE-3}
\end{equation}
The solution of the latter system is
\begin{equation}
\begin{aligned}
 p&=A\cos(\kappa (t-t_0))\;,\\
 q&=A\sin(\kappa (t-t_0))-\frac{H_0}{2\kappa}\;.
\end{aligned}
\label{apWE-4}
\end{equation}
Transforming back to $u,v$ we obtain
\begin{equation}
\begin{aligned}
 u=\rho\cos\Delta&=x_0+
 \frac{A\sin(\kappa
(t-t_0))-\frac{H_0}{2\kappa}}
{A^2+\frac{H_0^2}{4\kappa^2}-
\frac{AH_0}{\kappa\sin(\kappa
(t-t_0))}}\;, \\[2ex]
 v=\rho\sin\Delta&=
 \frac{A\cos(\kappa
(t-t_0))}
{A^2+\frac{H_0^2}{4\kappa^2}-
\frac{AH_0}{\kappa\sin(\kappa
(t-t_0))}}\;.
\end{aligned}
\label{apWE-5}
\end{equation}
In these variables Eq.~\reff{ws2-3} takes the form
\be
\dot\alpha=\w-\nu-H_0 u\;
\ee{apWE-6}
and is readily solved by substitution of \reff{apWE-5} and integration. 
As a result we obtain a quasiperiodic solution: variables $\rho$ and $\Delta$ 
oscillate with the frequency $\kappa$ (it means that $\Phi$ has the frequency $\kappa+\nu$) 
and $\Psi=\Phi-\alpha$
possesses an additional frequency, found via integration of \reff{apWE-6}.

\section{Variable transformation for continuity equation}
\label{AppCE}

We perform transformation of variables in Eq.~(\ref{conteq}), using Eq.~(\ref{eqdensnew}):
\begin{equation}
 \begin{array}{ll}
 0&=\fracpd{w}{t}+\frac{\pd}{\pd \phi}(w v) =\fracpd{w}{\tau}\fracpd{\tau}{t}
+\fracpd{w}{\psi}\fracpd{\psi}{t}+\frac{\pd}{\pd \psi}(w v)\frac{\pd\psi}{\pd\phi}\\[4ex]
&=\fracpd{}{\tau}\left(\sigma \fracpd{\psi}{\phi} \right)
+\fracpd{}{\psi}  \left(\sigma \fracpd{\psi}{\phi} \right)\cdot \fracpd{\psi}{t}+
\left[ \frac{\pd}{\pd \psi}\left(\sigma \fracpd{\psi}{\phi} \right)v+
 \left(\sigma \fracpd{\psi}{\phi} \right)\frac{\pd v}{\pd \psi} \right] \frac{\pd\psi}{\pd\phi}\\[4ex]
&=\fracpd{\sigma}{\tau}\fracpd{\psi}{\phi}+
\sigma\ds\frac{\pd}{\pd \tau}\left( \fracpd{\psi}{\phi}\right) 
+\left[ \fracpd{\sigma}{\psi}\fracpd{\psi}{\phi}+\sigma\ds\frac{\pd}{\pd \psi} 
\left( \fracpd{\psi}{\phi}\right)  \right] \fracpd{\psi}{t}\\[4ex]
&+\left\lbrace \left[\fracpd{\sigma}{\psi} \fracpd{\psi}{\phi}+\sigma\ds\frac{\pd}{\pd \psi} 
\left( \fracpd{\psi}{\phi}\right) \right] v+\sigma \fracpd{\psi}{\phi}\fracpd{v}{\psi}
\right\rbrace \fracpd{\psi}{\phi}\\[4ex]
&=\fracpd{\sigma}{\tau}\fracpd{\psi}{\phi}+
\sigma\left\lbrace \ds\frac{\pd}{\pd \tau}\left( \fracpd{\psi}{\phi}\right) 
+\ds\frac{\pd}{\pd \psi} \left( \fracpd{\psi}{\phi}\right)
\left (\fracpd{\psi}{t}+v\fracpd{\psi}{\phi}\right )+\left (\fracpd{\psi}{\phi}\right) ^2\fracpd{v}{\psi}
\right\rbrace\\[4ex]
&+\fracpd{\sigma}{\psi}\left\lbrace 
\fracpd{\psi}{\phi}\left (\fracpd{\psi}{t}+v\fracpd{\psi}{\phi}\right)\right\rbrace\;.
\end{array}
\label{cenew1}
\end{equation}

Let us demonstrate that the coefficients at $\sigma$ and $\fracpd{\sigma}{\psi}$ vanish 
if $\rho$, $\Phi$, and $\Psi$ obey the WS equations. For this goal we first compute 
$\fracpd{\psi}{t}+v\fracpd{\psi}{\phi}$. 
It is convenient to use the notations $f=e^{i(\phi-\Phi)}$,  $c=e^{i(\psi-\Psi)}$.
Resolving Eq.~(\ref{ws-trans}) with respect 
to $\psi$, we obtain
\begin{equation}
 \psi=\Psi-i\ln(f-\rho)+i\ln(1-\rho f)\;.
\label{psiphi}
\end{equation}
Taking the derivative and re-arranging the terms, we obtain
\begin{equation}
 \frac{\pd\psi}{\pd t}(\phi)=\dot\Psi-f\frac{1-\rho^2}{(f-\rho)(1-f\rho)} \dot\Phi
+i\frac{1-f^2}{(f-\rho)(1-f\rho)}\dot\rho\;.
\label{psi_t}
\end{equation}
Using $e^{-i\phi}=e^{-i\Phi}/f=e^{-i\Phi}f^*$, we obtain in new variables:
\begin{equation}
v=\w+\mbox{Im}\left [He^{-i\Phi}f^*\right] \;.
\label{app2:1}
\end{equation}
Next, from Eq.~(\ref{psiphi}) we compute, using $\fracpd{f}{\phi}=if$:
\begin{equation}
 \frac{\pd\psi}{\pd\phi}(\phi)=\frac{(1-\rho^2)f}{(f-\rho)(1-\rho f)} \;.
\label{app2:2}
\end{equation}
Substituting into Eq.~(\ref{psi_t}) the derivatives via the r.h.s. of the WS equations and 
using Eqs.~(\ref{app2:1},\ref{app2:2}), we obtain after tedious but straightforward algebra
\begin{equation}
 \fracpd{\psi}{t}+v\fracpd{\psi}{\phi}=0\;.
\end{equation}

Hence the coefficient at $\fracpd{\sigma}{\psi}=0$ and the coefficient at $\sigma$ reduces to 
\[
\fracpd{}{\tau}\left (\fracpd{\psi}{\phi}\right )+\left (\fracpd{\psi}{\phi}\right) ^2\fracpd{v}{\psi}=Q\;.
\]
To compute $Q$, we first substitute in Eq.~(\ref{app2:2}) $f=\ds\frac{\rho + c}{\rho c +1}$ 
from Eq.~(\ref{ws-trans}) and obtain, after straightforward manipulations,
\begin{equation}
 \frac{\pd\psi}{\pd\phi}(\psi)=\frac{(\rho +c)(\rho+ c^*)}{1-\rho^2}
=\frac{\rho c+ \rho c^*+2}{1-\rho^2}-1\;.
\label{app2:3}
\end{equation}
Derivation with respect to time yields
\begin{equation}
 \ds\frac{\pd}{\pd \tau}\left( \fracpd{\psi}{\phi}\right)=
\ds\frac{\pd}{\pd t}\left( \fracpd{\psi}{\phi}\right)=\frac{i\rho(c^*-c)}{1-\rho^2}\dot\Psi+
\frac{(1+\rho^2)(c+c^*)+4\rho}{(1-\rho^2)^2}\dot\rho\;.
\label{app2:4}
\end{equation}
Here we used $\fracpd{c}{t}=-ic\dot\Psi$,  $\fracpd{c^*}{t}=ic^*\dot\Psi$.
Next, we compute 
\begin{equation}
 \fracpd{v}{\psi}=\mbox{Im}\left[He^{-i\Phi} \ds\frac{\pd}{\pd\psi}\frac{\rho c +1}{\rho+c}  
\right]=(\rho^2-1)\mbox{Re}\left[\ds\frac{cHe^{-i\Phi}}{(\rho+c)^2}\right] \;.
\label{app2:5}
\end{equation}
Using the obtained expressions (\ref{app2:3}-\ref{app2:5}), we show, after tedious but 
straightforward manipulations, that $Q=0$ if $\dot\Psi$ and $\dot\rho$ obey the WS equations.

Thus, we demonstrate that the r.h.s. of the continuity equation Eq.~(\ref{cenew1}) simplifies to 
$\fracpd{\sigma}{\tau}\fracpd{\psi}{\phi}$ and is therefore valid if
$\sigma(\w,\psi)$ is a stationary 
distribution.

\section{Choice of initial conditions for simulation of hierarchical populations}
\label{AppIC}

Our goal is to choose different microscopic initial conditions, i.e. initial values for oscillator 
phases, but keep the same  macroscopic initial conditions, i.e. the amplitude of the mean field.
For this goal we proceed as follows.
For each subpopulation with the frequency 
$\w_a$ we take $\psi_a$ uniformly distributed 
along the arcs $[(1-q)\frac{\pi}{2},(1+q)\frac{\pi}{2}]$ and  
$[-(1+q)\frac{\pi}{2}],-(1-q)\frac{\pi}{2}]$, 
as shown in Fig.~\ref{figpsi}. 
Here $0<q\le 1$ is a parameter quantifying deviation of the distribution from a 
uniform one; $q=1$ corresponds do a uniform distribution, with $q\to 0$ the 
distribution collapses to two points.
For this construction, the subpopulation $N_a$ should be an even number.
Note that this choice of $\psi_{a,k}$ satisfies constraints (\ref{psicond}) and $\sum\psi_{a,k}=0$.
The initial values of the oscillator phases $\phi_{a,k}(0)$ are obtained from $\psi_{a,k}$ according 
to Eq.~(\ref{ap1-7}).

\begin{figure}[ht!]
\centerline{\includegraphics[width=0.4\textwidth]{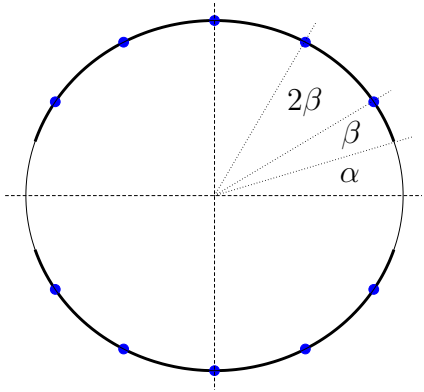}}
\caption{Illustration of the special choice of the constants of motion $\psi_k$, here for 
$q=0.8$ and $N_a=10$. The points are distributed along two arcs of length $q\pi$ each;
angle $\alpha=\ds\frac{\pi}{2}(1-q)$, angle $\beta=\ds\frac{q\pi}{N_a}$.
}
\label{figpsi}
\end{figure}

Now we show that with a special choice of the initial values of the WS variables 
we can ensure the same initial value of the mean field, independently of the 
parameter $q$.
These special values are $\Phi_a=0$, $\rho_a=\rho_0$, and $\Psi_a=\frac{2\pi}{M}a$. 
In order to compute the initial value of the Kuramoto mean field 
$Y(0)=r_0e^{i\Theta_0}$ we write the discrete version of Eq.~(\ref{gop-1}) for $t=0$:
\begin{align*}
 r_0e^{i\Theta_0}&=\rho_0\sum_{a=1}^M n_a \gamma_a\\[2ex]
&=\rho_0\left (\sum_{a=1}^M n_a +
(1-\rho_0^{-2})\sum_{a=1}^M n_a\sum_{l=2}^\infty C_{l}(-\rho_0)^l e^{-i\frac{2\pi al}{M}}
\right )\\[2ex]
&=\rho_0\;,
\end{align*}
with account that all groups are of equal size, $n_a=n$.
Thus, taking different values of the parameter $q$ and fixing other parameters we obtain 
the same macroscopic initial conditions (i.e. for the mean field), whereas the initial conditions 
for individual oscillators are different.


\begin{thebibliography}{10}

\bibitem{Wiesenfeld-Swift-95}
K.~Wiesenfeld and J.~W. Swift.
\newblock Averaged equations for {J}osephson junction series arrays.
\newblock {\em Phys. Rev. E}, 51(2):1020--1025, 1995.

\bibitem{Glova-03}
A.~F. Glova.
\newblock Phase locking of optically coupled lasers.
\newblock {\em Quantum Electronics}, 33(4):283--306, 2003.

\bibitem{Kiss-Zhai-Hudson-02a}
I.Z. Kiss, Y.~Zhai, and J.L. Hudson.
\newblock Emerging coherence in a population of chemical oscillators.
\newblock {\em Science}, 296:1676--1678, 2002.

\bibitem{Strogatz_et_al-05}
S.~H. Strogatz, D.~M. Abrams, A.~McRobie, B.~Eckhardt, and E.~Ott.
\newblock Theoretical mechanics: {C}rowd synchrony on the {M}illennium
  {B}ridge.
\newblock {\em Nature}, 438:43--44, 2005.

\bibitem{Eckhardt_et_al-07}
B.~Eckhardt, E.~Ott, S.~H. Strogatz, D.~M. Abrams, and A.~McRobie.
\newblock Modeling walker synchronization on the {M}illennium {B}ridge.
\newblock {\em Phys. Rev. E}, 75:021110, 2007.

\bibitem{Neda-Ravasz-Brechet-Vicsek-Barabasi-00}
Z.~N{\'e}da, E.~Ravasz, Y.~Brechet, T.~Vicsek, and A.-L. Barab{\'a}si.
\newblock Tumultuous applause can transform itself into waves of synchronized
  clapping.
\newblock {\em Nature}, 403(6772):849--850, 2000.

\bibitem{Richard-Bakker-Teusink-Van-Dam-Westerhoff-96}
P.~Richard, B.~M. Bakker, B.~Teusink, K.~Van Dam, and H.~V. Westerhoff.
\newblock Acetaldehyde mediates the synchronization of sustained glycolytic
  oscillations in population of yeast cells.
\newblock {\em Eur. J. Biochem.}, 235:238--241, 1996.

\bibitem{Dano-Sorensen-Hynne-99}
S.~Dano, P.~G. Sorensen, and F.~Hynne.
\newblock Sustained oscillations in living cells.
\newblock {\em Nature}, 402(6759):320--322, 1999.

\bibitem{Gonze-Markadieu-Goldbeter-08}
D.~Gonze, N.~Markadieu, and A.~Goldbeter.
\newblock {Selection of in-phase or out-of-phase synchronization in a model
  based on global coupling of cells undergoing metabolic oscillations}.
\newblock {\em {CHAOS}}, {18}({3}):{037127}, {SEP} {2008}.

\bibitem{Golomb-Hansel-Mato-01}
D.~Golomb, D.~Hansel, and G.~Mato.
\newblock Mechanisms of synchrony of neural activity in large networks.
\newblock In F.~Moss and S.~Gielen, editors, {\em Neuro-informatics and Neural
  Modeling}, volume~4 of {\em Handbook of Biological Physics}, pages 887--968.
  Elsevier, Amsterdam, 2001.

\bibitem{Sakaguchi-88}
H.~Sakaguchi.
\newblock Cooperative phenomena in coupled oscillator systems under external
  fields.
\newblock {\em Prog. Theor. Phys.}, 79(1):39--46, 1988.

\bibitem{Antonsen-Faghih-Girvan-Ott-08}
T.~M. Antonsen, Jr., R.~T. Faghih, M.~Girvan, E.~Ott, and J.~Platig.
\newblock External periodic driving of large systems of globally coupled phase
  oscillators.
\newblock {\em Chaos}, 18(3):037112, 2008.

\bibitem{Tass-99}
P.~A. Tass.
\newblock {\em Phase Resetting in Medicine and Biology. Stochastic Modelling
  and Data Analysis.}
\newblock Springer-Verlag, Berlin, 1999.

\bibitem{Rosenblum-Pikovsky-04b}
M.~Rosenblum and A.~Pikovsky.
\newblock Controlling synchronization in an ensemble of globally coupled
  oscillators.
\newblock {\em Phys. Rev. Lett.}, 92(11):114102, 2004.

\bibitem{Rosenblum-Pikovsky-04c}
M.~Rosenblum and A.~Pikovsky.
\newblock Delayed feedback control of collective synchrony: {A}n approach to
  suppression of pathological brain rhythms.
\newblock {\em Phys. Rev. E}, 70:041904, 2004.

\bibitem{Childs-Strogatz-08}
L.~M. Childs and S.~H. Strogatz.
\newblock Stability diagram for the forced {K}uramoto model.
\newblock {\em Chaos: An Interdisciplinary Journal of Nonlinear Science},
  18(4):043128, 2008.

\bibitem{Martens_etal-09}
E.~A. Martens, E.~Barreto, S.~H. Strogatz, E.~Ott, P.~So, and T.~M. Antonsen.
\newblock Exact results for the {K}uramoto model with a bimodal frequency
  distribution.
\newblock {\em Physical Review E}, 79(2):026204, 2009.

\bibitem{Abrams-Mirollo-Strogatz-Wiley-08}
D.~M. Abrams, R.~Mirollo, S.~H. Strogatz, and D.~A. Wiley.
\newblock Solvable model for chimera states of coupled oscillators.
\newblock {\em Phys. Rev. Lett.}, 101:084103, 2008.

\bibitem{Omelchenko-Maistrenko-Tass-08}
O.~E. Omel'chenko, Yu.~L. Maistrenko, and P.~A. Tass.
\newblock Chimera states: The natural link between coherence and incoherence.
\newblock {\em Physical Review Letters}, 100(4):044105, 2008.

\bibitem{Laing-09}
C.~R. Laing.
\newblock The dynamics of chimera states in heterogeneous {K}uramoto networks.
\newblock {\em Physica D: Nonlinear Phenomena}, 238(16):1569 -- 1588, 2009.

\bibitem{Rosenblum-Pikovsky-07}
M.~Rosenblum and A.~Pikovsky.
\newblock Self-organized quasiperiodicity in oscillator ensembles with global
  nonlinear coupling.
\newblock {\em Phys. Rev. Lett.}, 98:064101, 2007.

\bibitem{Pikovsky-Rosenblum-09}
A.~Pikovsky and M.~Rosenblum.
\newblock Self-organized partially synchronous dynamics in populations of
  nonlinearly coupled oscillators.
\newblock {\em Physica D}, 238(1):27--37, 2009.

\bibitem{Ott-Antonsen-08}
E.~Ott and Th.~M. Antonsen.
\newblock Low dimensional behavior of large systems of globally coupled
  oscillators.
\newblock {\em CHAOS}, 18(3):037113, 2008.

\bibitem{Ott-Antonsen-09}
E.~Ott and Th.~M. Antonsen.
\newblock {Long time evolution of phase oscillator systems}.
\newblock {\em {CHAOS}}, {19}({2}):{023117}, {2009}.

\bibitem{Pazo-05}
D.~Pa{\'z}o.
\newblock {Thermodynamic limit of the first-order phase transition in the
  {K}uramoto model}.
\newblock {\em {Phys. Rev. E}}, {72}({4, Part 2}):{046211}, {OCT} {2005}.

\bibitem{Pazo-Montbrio-09}
D.~Pa{\'z}o and E.~Montbri{\'o}.
\newblock Existence of hysteresis in the {K}uramoto model with bimodal
  frequency distributions, 2009.
\newblock {\em Phys. Rev. E}, 80: 046215, 2009.

\bibitem{Kori-Kuramoto-01}
H.~Kori and Y.~Kuramoto.
\newblock Slow switching in globally coupled oscillators: robustness and
  occurrence through delayed coupling.
\newblock {\em Phys. Rev. E}, 63:046214, 2001.

\bibitem{Liu-Lai-Hoppensteadt-01}
Z.~Liu, Y.-C. Lai, and F.~C. Hoppensteadt.
\newblock Phase clustering and transition to phase synchronization in a large
  number of coupled nonlinear oscillators.
\newblock {\em Phys. Rev. E}, 63(5):055201, 2001.

\bibitem{Maistrenko-Popovych-Burylko-Tass-04}
Yu. Maistrenko, O.~Popovych, O.~Burylko, and P.~A. Tass.
\newblock Mechanism of desynchronization in the finite-dimensional {K}uramoto
  model.
\newblock {\em Phys. Rev. Lett.}, 93(8):084102, Aug 2004.

\bibitem{Kuramoto-75}
Y.~Kuramoto.
\newblock Self-entrainment of a population of coupled nonlinear oscillators.
\newblock In H.~Araki, editor, {\em International Symposium on Mathematical
  Problems in Theoretical Physics}, page 420, New York, 1975. Springer Lecture
  Notes Phys., v. 39.

\bibitem{Kuramoto-84}
Y.~Kuramoto.
\newblock {\em Chemical Oscillations, Waves and Turbulence}.
\newblock Springer, Berlin, 1984.

\bibitem{Daido-92a}
H.~Daido.
\newblock Order function and macroscopic mutual entrainment in uniformly
  coupled limit-cycle oscillators.
\newblock {\em Prog. Theor. Phys.}, 88(6):1213--1218, 1992.

\bibitem{Daido-93a}
H.~Daido.
\newblock Critical conditions of macroscopic mutual entrainment in uniformly
  coupled limit-cycle oscillators.
\newblock {\em Prog. Theor. Phys.}, 89(4):929--934, 1993.

\bibitem{Daido-96}
H.~Daido.
\newblock Onset of cooperative entrainment in limit-cycle oscillators with
  uniform all-to-all interactions: {B}ifurcation of the order function.
\newblock {\em Physica D}, 91:24--66, 1996.

\bibitem{Pikovsky-Rosenblum-Kurths-01}
A.~Pikovsky, M.~Rosenblum, and J.~Kurths.
\newblock {\em Synchronization. A Universal Concept in Nonlinear Sciences.}
\newblock Cambridge University Press, Cambridge, 2001.

\bibitem{Acebron-etal-05}
J.~A. Acebron, L.~L. Bonilla, C.~J.~Perez Vicente, F.~Ritort, and R.~Spigler.
\newblock The {K}uramoto model: {A} simple paradigm for synchronization
  phenomena.
\newblock {\em Rev. Mod. Phys.}, 77(1):137--175, 2005.

\bibitem{Strogatz-00}
S.~H. Strogatz.
\newblock From {K}uramoto to {C}rawford: {E}xploring the onset of
  synchronization in populations of coupled oscillators.
\newblock {\em Physica D}, 143(1-4):1--20, 2000.

\bibitem{Watanabe-Strogatz-93}
S.~Watanabe and S.~H. Strogatz.
\newblock Integrability of a globally coupled oscillator array.
\newblock {\em Phys. Rev. Lett.}, 70(16):2391--2394, 1993.

\bibitem{Watanabe-Strogatz-94}
S.~Watanabe and S.~H. Strogatz.
\newblock Constants of motion for superconducting {J}osephson arrays.
\newblock {\em Physica D}, 74:197--253, 1994.

\bibitem{Pikovsky-Rosenblum-08}
A.~Pikovsky and M.~Rosenblum.
\newblock Partially integrable dynamics of hierarchical populations of coupled
  oscillators.
\newblock {\em Phys. Rev. Lett.}, 101:264103, 2008.

\bibitem{Marvel-Mirollo-Strogatz-09}
S.~A. Marvel, R.~E. Mirollo, and S.~H. Strogatz.
\newblock Phase oscillators with global sinusoidal coupling evolve by {M}obius
  group action.
\newblock {\em arXiv:0904.1680}, 2009.
\newblock (CHAOS, submitted).

\bibitem{Barreto-Hunt-Ott-So-08}
E.~Barreto, B.~Hunt, E.~Ott, and P.~So.
\newblock Synchronization in networks of networks: {T}he onset of coherent
  collective behavior in systems of interacting populations of heterogeneous
  oscillators.
\newblock {\em Phys. Rev. E}, 77:036107, 2008.

\bibitem{Lee-Ott-Antonsen-09}
W.~S.~Lee, E.~Ott, Th.~M.~Antonsen.
\newblock Large coupled oscillator systems with heterogeneous interaction delays.
\newblock {\em Phys. Rev. Lett.}, 103:044101, 2009.

\bibitem{Abdulrehem-Ott-09}
M.~M.~Abdulrehem, E.~Ott.
\newblock Low dimensional description of pedestrian-induced oscillation of 
the Millennium Bridge.
\newblock {\em {CHAOS}}, {19}({1}):{013129}, {2009}.

\bibitem{Daido-95}
H.~Daido.
\newblock Multi-branch entrainment and multi-peaked order-functions in a phase
  model of limit-cycle oscillators with uniform all-to-all coupling.
\newblock {\em J. Phys. A: Math. Gen.}, 28:L151--L157, 1995.

\bibitem{Sakaguchi-Kuramoto-86}
H.~Sakaguchi and Y.~Kuramoto.
\newblock A soluble active rotator model showing phase transition via mutual
  entrainment.
\newblock {\em Prog. Theor. Phys.}, 76(3):576--581, 1986.

\bibitem{Ott-Platig-Antonsen-Girvan-08}
E.~Ott, J.~H. Platig, Th.~M.~Antonsen, and M.~Girvan.
\newblock Echo phenomena in large systems of coupled oscillators.
\newblock {\em {CHAOS}}, {18}({3}):{037115}, {2008}.

\end{thebibliography}

\end{document}